\documentclass[12pt,preprint]{aastex}
\usepackage{epsfig}
\usepackage{psfig}

\newcommand{\eo}{{\sl Einstein Observatory}}
\newcommand{\asca}{{\sl ASCA}}
\newcommand{\ASCA}{{\sl ASCA}}
\newcommand{\Einstein}{{\sl Einstein}}
\newcommand{\xmm}{{\sl XMM-Newton}}
\newcommand{\XMM}{{\sl XMM-Newton}}

\newcommand{\rosat}{{\sl ROSAT}}
\newcommand{\ROSAT}{{\sl ROSAT}}

\newcommand{\cha}{{\sl Chandra}}

\slugcomment{ to be submitted}
\shorttitle{{Six Years of \cha\ Observations of SNRs}}
\shortauthors{}

\begin{document}

\title{Six Years of \cha\ Observations of Supernova Remnants}

\author{
Martin C. Weisskopf \altaffilmark{1} \&
John P. Hughes \altaffilmark{2}
}

\altaffiltext{1}
{Space Sciences Directorate, NASA Marshall Space Flight Center, XD01, Huntsville, AL 35812}
\altaffiltext{2}
{Department of Physics and Astronomy, Rutgers University, 136 Frelinghusyen Road, Piscataway, NJ 08854}

\begin{abstract}

We present a review of the first six years of \cha\ X-ray Observatory
observations of supernova remnants.  From the official "first-light"
observation of Cassiopeia A that revealed for the first time the
compact remnant of the explosion, to the recent million-second
spectrally-resolved observation that revealed new details of the
stellar composition and dynamics of the original explosion, \cha\
observations have provided new insights into the supernova
phenomenon.   We present an admittedly biased overview of six
years of these observations, highlighting new discoveries made
possible by \cha's unique capabilities.

\end{abstract}


\section{Introduction\label{s:intro}}

The \cha\ X-ray Observatory was launched on July 23, 1999 using the Space Shuttle Columbia. 
Placement in its highly elliptical orbit was completed 15 days after launch.  
The orbit has a nominal apogee of 140,000 km and a nominal perigee of 10,000 km. 
With this orbit, the satellite is above the radiation belts for more than about
75\% of the 63.5-hour orbital period and uninterrupted observations lasting more than 2 days are possible.
The observing efficiency, which also depends on solar activity, is typically about 65\%.

The spacecraft has been functioning superbly since launch.  
The specified design life of the mission was 5 years; however, the only
perishable is gas for maneuvering and is sized to allow operation for much more than 10 years. 
The orbit will be stable for decades.

The heart of the Observatory is the X-ray telescope made of four concentric,
precision-figured, superpolished Wolter-1 telescopes, similar to those
used for both the \Einstein\ and \ROSAT\ observatories, but of much higher quality, larger diameter, and longer focal length. 

The telescope's on-axis point spread function, as measured during ground calibration, had a full-width at half-maximum less than 0.5 arcsec and a half-power diameter less than 1 arcsec. 
A relatively mild dependence on energy, resulting from diffractive scattering by
surface microroughness, attests to the better than 3-angstroms-rms surface
roughness measured with optical metrology during fabrication and confirmed by
the ground X-ray testing.
The on-orbit performance met expectations.

There are two focal plane cameras on the Observatory. 
The Pennsylvania State University (PSU, University Park, Pennsylvania) and the Massachusetts Institute of Technology (MIT, Cambridge, Massachusetts) designed and fabricated the Advanced CCD Imaging Spectrometer (ACIS) with CCDs produced by MIT's Lincoln Laboratory. 
Made of a 2-by-2 array of front-illuminated (FI), 2.5-cm-square
CCDs, ACIS-I provides high-resolution spectrometric imaging over a
17-arcmin-square field of view.
ACIS-S, a 6-by-1 array of 4 FI CCDs and two back-illuminated (BI) CCDs mounted
along the \cha\ transmission grating dispersion direction, serves both as the primary read-out detector for the High Energy Transmission Grating (HETG),
and, using the one BI CCD which can be placed at the aim point of the telescope,
also provides high-resolution spectrometric imaging extending to lower energies
but over a smaller (8-arcmin-square) field than ACIS-I.

The Smithsonian Astrophysical Observatory (SAO, Cambridge, Massachusetts), designed and fabricated the other focal plane camera the High Resolution Imager (HRC) (Murray et al.\ 2000).
Made of a single 10-cm-square microchannel plate (MCP), the HRC-I provides
high-resolution imaging over a 30-arcmin-square field of view.
A second detector made of 3 rectangular MCP segments (3-cm $\times$ 10-cm each) mounted end-to-end along the grating dispersion direction, the HRC-S, serves as the primary read-out detector for the Low Energy Transmission Grating (LETG).

More details as to the technical performance of the Observatory may be found in
Weisskopf et al.\ (2003).

We note that the names of SNR are often given in papers using their galactic coordinates and a ''G'' prefix. 
This prefix is, however, used by name resolvers such as Simbad for high proper motion stars appearing in the Giclas Catalog and it has been recommended \footnote{see http://heasarc.gsfc.nasa.gov/docs/faq.html\#nameresolvers} that one use the SNR designation instead. 
We shall do so in this review.

Our review is representative of \cha\ observations and is not intended to be exhaustive.
Thus, if we have omitted a particular observation no slight to the work was intended.
Finally, we emphasize that there are many excellent reviews and conference proceedings concerning the astrophysics associated with supernova remnants and, where relevant, the compact objects within them. 
These reviews and proceedings include: Becker and Pavlov (2001), Aschenbach (2002) Pavlov, Zavlin, and Sanwal (2002), Slane (2002), Canizares (2004), Decourchelle (2004), Slane (2005), and Rakowski (2005).

Our purpose here is to review the outstanding contributions that observations using \cha\ have made to the discipline. 
We urge interested readers to include a careful reading of these reviews in addition to reading this paper. 
 
\section{The Point Sources in Supernova Remnants\label{s:pt_sources}}

An excellent introduction to the topic of compact central sources in supernova remnants may be found in Kaplan et al.\ (2004). 

\subsection{The Compact Central Objects\label{ss:cco}}

Although not originally discovered with \cha, observations using the
observatory have both clarified, and drawn further attention to, the
existence of a group of compact objects associated with SNR that are
radio quiet, unpulsed in X-rays, and have characteristic X-ray spectra
described by a blackbody with characteristic temperatures of order 0.4
keV, without indication of a non-thermal component.  The associated
blackbody radii are smaller than the radii of canonical neutron stars.
These objects have been aptly termed Compact Central Objects (CCOs) by
Pavlov et al.\ (2002a).  (Sources of this ilk have previously been
referred to as ``Radio Quiet Neutron Stars'' by Caraveo, Bignami, \&
Tr\"umper 1996.)  Although several sources exhibit these properties,
it is not clear that they form a class in the sense that there is
an universally accepted explanation for their characteristics.

The prevailing, perhaps contradictory, interpretation is that the
spectral properties of the CCOs are the result of hot spots on the
surfaces of neutron stars formed during the supernova.  The inability
then to find any evidence for pulsations in the X-ray flux from the
majority of these sources must be confronted.  Many discussions leave
one with the impression that new observations and more sensitive
searches for pulsations will ultimately discover periodic behavior and
thus validate the explanation.  In general, detection of pulsations
has not been forthcoming, and so perhaps deeper looks at other
potential explanations might be in order.  One possible explanation
involves very long periods requiring some mechanism to carry off the
angular momentum.  One might also begin to question the interpretation
of the spectral data.  The tie between spectral fits and physical
interpretations can be overdone.  Thus, simply because a blackbody
spectrum fits the data, this does not necessarily imply that the
source is a blackbody emitter.  Another spectrum, more representative
of the true physical situation, may fit the data equally well.  It is
a challenge to the theorists to tell us just what this alternative
spectrum might be.

\subsubsection{Cas A\label{sss:casa}}

\begin{figure}
\begin{center} 
\resizebox{4in}{!}{\rotatebox{0}{\includegraphics{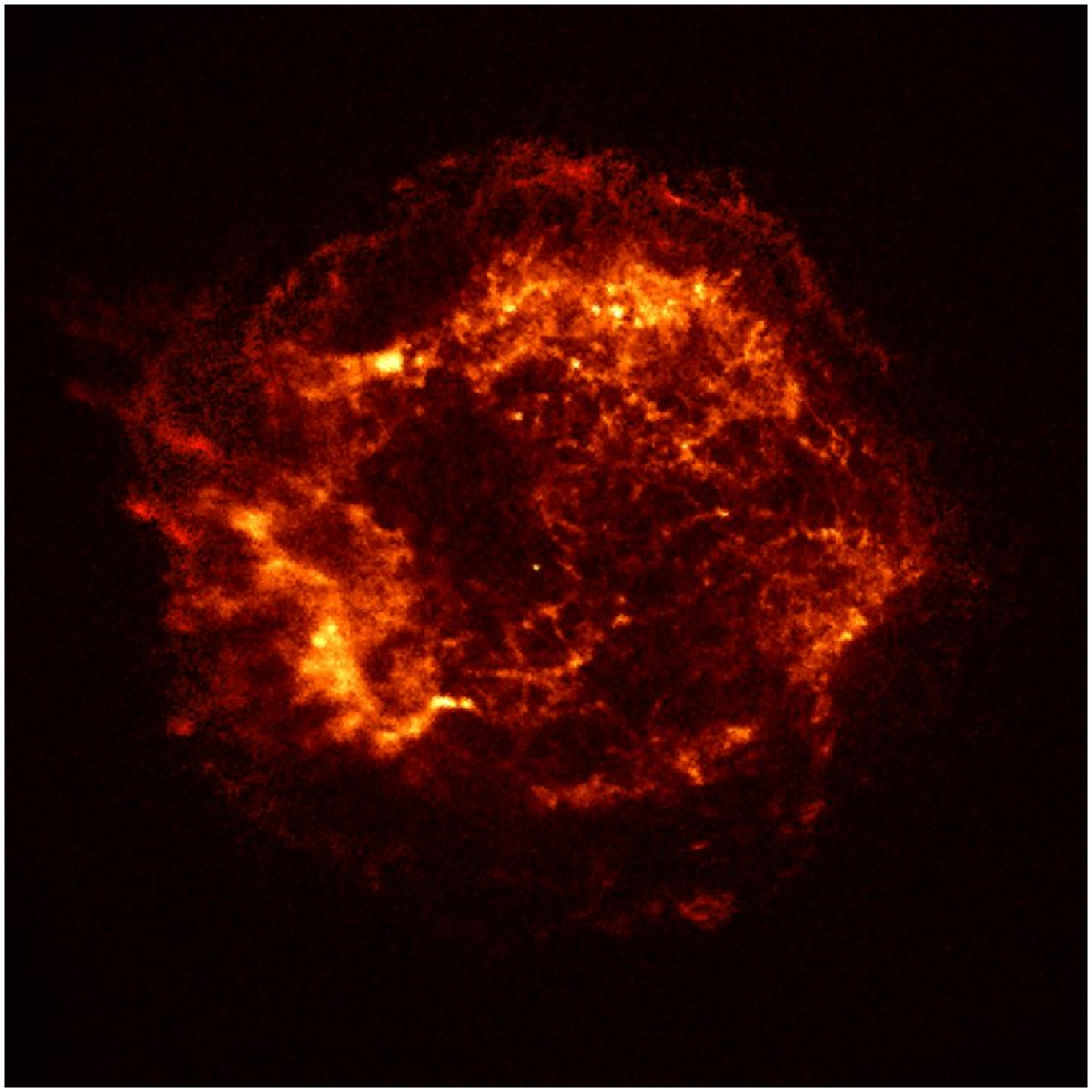}}}
\caption{\label{f:CasA} Early \cha\ ACIS-S3 image of Cas A. The image is 6' x 6'. Courtesy NASA/CXC/SAO.
}
\end{center}
\end{figure}

Cassiopeia A (Cas A) is often referred to as the ``first light''
observation made with \cha.  It is historically interesting that this
is not quite accurate.  The first X-rays that were observed with \cha\
when the last door was opened took place on August 12-th, 1999 and led
to the discovery of the z=0.32 AGN nicknamed ``Leon X-1'' (Weisskopf
et al.\ 2005).  The nickname honors the \cha\ Telescope Scientist,
Leon Van Speybroeck.  The Observatory was next pointed at the
radio-loud quasar PKS 0637$-$752 (Schwartz et al.\ 2000) chosen as the
target for which on-orbit optimization of the imaging properties of
the Observatory were performed.  After bore-sighting and focusing the
telescope, only then was the official ``first-light'' image of Cas A
obtained.  One of the principal features of the first Cas A image,
shown in Figure~\ref{f:CasA}, was the clear presence of a point source
(Tananbaum 1999) near the center of the remnant and apparent after
only a few minutes of observing.

Pavlov et al.\ (2000) argued, convincingly, that the source was the
compact object associated with the remnant.  Pavlov et al.\ (2000) also
established the nature of the spectral parameters, noting that no
unique spectral fits were forthcoming from the data in hand at the
time. Fits to a power law yielded high photon indices (of order 3 to
4) whereas fits to a blackbody led to temperatures of about 0.5 keV.
As noted above, these authors coined the very apt description for this
source --- Compact Central Object (CCO) --- and favored the
interpretation that the CCO is a neutron star whose X-ray emission is
primarily from a restricted region of the surface.

Subsequently, Chakrabarty et al.\ (2001) provided additional analyses
of both spectral and timing data, setting a 3-$\sigma$ upper limit to
the pulsed fraction of less than 35\% for periods longer than 20-ms.
Additional period searches were performed by Murray et al.\ (2002a)
utilizing HRC-S data in order to extend the period searches to shorter
periods.  Using a 50 ksec observation they also failed to detect any
significant evidence for pulsations.  More recently, and discussed in
detail in [\S~\ref{jph:casa}] a $10^{6}$ s
ACIS-S3 observation has been performed.  To our knowledge, no
pulsations have thus far been detected in these data, although the
search would be limited to periods longer than several seconds due to the time resolution in this mode.

\subsubsection{SNR 266.2$-$1.2 (RX J09852.0$-$4622)\label{sss:SNR266.2-1.2}}

The SNR 266.2$-$1.2 (aka RX J09852.0$-$4622 and sometimes referred
to as ``Vela Junior'') was discovered with \ROSAT\ by Aschenbach
(1998).  Prior to observations with \cha, a number of X-ray
observations had been performed with at least two, if not more,
possible X-ray point sources being associated with this remnant -- see
the introduction to, and references in, Pavlov et al.\ (2001).  Pavlov
et al.\ (2001) used \cha\ and the ACIS imaging array to pinpoint the
location of a bright X-ray source (CXOU J085201.4$-$461753) 4$^\prime$
north of the SNR center.  This was the only bright X-ray source they
found near the center of the remnant.  Using the \cha-based position
to refine and guide a search for optical counterparts, they found none
and established stringent upper limits to the optical flux in both B
and R.  These authors then used the corresponding lower limit to the
X-ray to optical flux ratio to argue that the source is the compact
remnant of the SN.  The quality of the X-ray spectra determined from
this particular observation was limited by the short observing time (3
ksec) and by pulse pileup due to the brightness of the source coupled
with the readout time (frame time) of the CCDs.

To refine the spectral parameters, and to provide moderately high time
resolution data useful for searching for pulsations, the source was
again observed with \cha\ by Kargaltsev et al.\ (2002) using ACIS-I in
the continuous-clocking mode.  This mode provides only one-dimensional
images but yields higher time resolution (2.85 ms in this case)
especially useful for mitigating pulse pileup.  This observation was
also ten times longer than the previous \cha\ observation.  Spectral
fits to a power law were found to be statistically unacceptable,
whereas a blackbody model, with a temperature of 0.4 keV, provided a
statistically acceptable fit.  As with Cas A and the other CCOs, the
radius of the emitting sphere was found to be much smaller than a
canonical neutron star radius.

Kargaltsev et al.\ (2002) also attempted to constrain the temperature
of the entire (putative) neutron star surface by fitting the spectral
data with a two-component blackbody model where one of the two
components was constrained to a radius of 10 km.  They found an upper
limit to the surface temperature (at infinity) of 89 eV (99\%
confidence) which would imply accelerated cooling compared to standard
neutron star cooling models assuming an age of a few thousand years.
(Inferring accelerated cooling is not exceptional, see, e.g. the discussion of 3C58 in \S~\ref{sss:3c58}.)  Kargaltsev et al.\ (2002) also searched for narrow
spectral features and discussed the hint of a feature at 1.68 keV.  It
would be interesting to follow up on the possible spectral feature
with \XMM\ to see if this is similar to the feature(s) discovered
using \cha\ in 1E 1207.4$-$5209 (\S~\ref{ss:SNR296.5+10}).  A search
for pulsations found no statistically significant periods with pulsed
fractions in excess of 13\% in the frequency range from 0.001 to 100
Hz.

We urge other observers to follow the example set by Kargaltsev et
al.\ (2002) in their data analysis.  In particular, use the data to
establish an upper limit to a possible full-surface blackbody in order
to constrain a neutron star temperature, and search for possible
spectral features that appear as residuals to the continuum models.
The results of these analyses are very useful in addressing some of
the potentially interesting astrophysical questions concerning compact
objects and in particular the CCOs.

\subsubsection{Puppis A\label{sss:puppisa}}

\begin{figure} 
\begin{center}
\resizebox{5in}{!}{\rotatebox{0}{\includegraphics{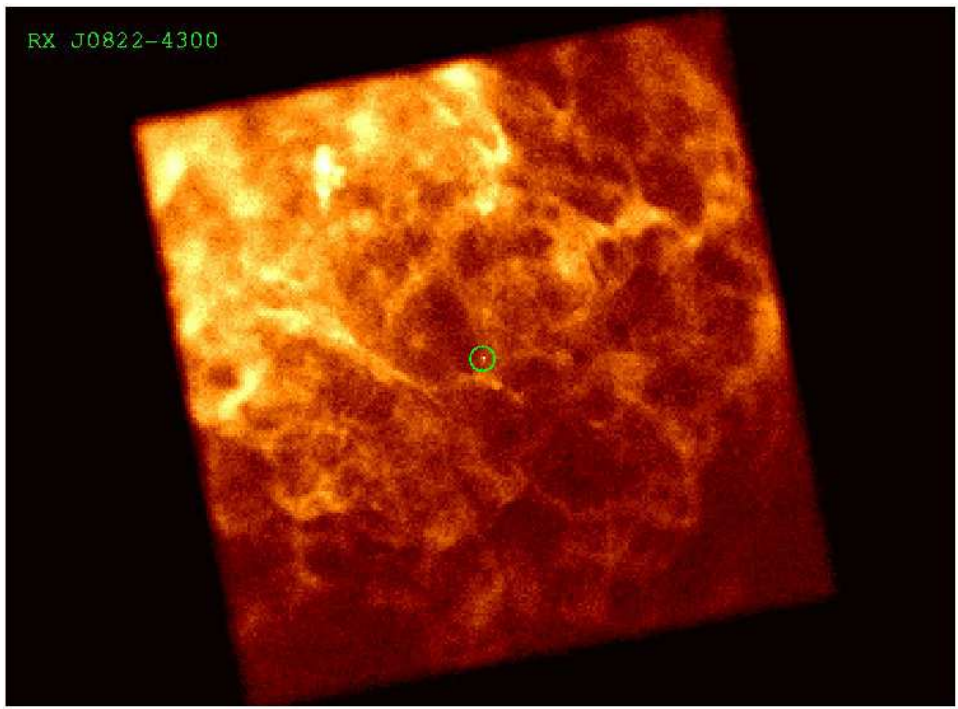}}}
\caption{\label{f:puppisa} \cha\ HRC-I image of Puppis-A. The image is 30' x 30'. Courtesy G. Pavlov.
}
\end{center}
\end{figure}

The bright point source at about 6$^\prime$ from the dynamical center of
Puppis A was discovered by Petre et al.\ (1982) using the \Einstein\
Observatory.  The object has been studied with numerous X-ray
satellites and in other wavelength bands (see e.g. the introduction to
Hui and Becker (2005) and references therein).  The early \cha\ ACIS
observation discussed by Pavlov et al.\ (2002a) found a spectrum both
consistent with previous X-ray observations and the CCO classification
--- namely that the data were described by a blackbody with an
associated radius much less than that of a neutron star.  In this
case, spectral fitting with a neutron star hydrogen atmosphere yielded
a radius more like that of a neutron star.  The ACIS observations were
followed by observations with the HRC designed to search both for
pulsations and a pulsar wind nebula (PWN).  The HRC image is shown in
Figure~\ref{f:puppisa} and clearly there is no obvious evidence for a
PWN, however Pavlov et al.\ (2002a) gave no quantitative upper limits.
Note that Gaensler, Bock and Stappers (2000) did provide quantitative
upper limits as to the presence of a radio PWN (on scales greater than
about 30" x 30 " up to about 30').  Pavlov et al.\ (2002a) used the
HRC-data to search for pulsations and found no evidence for pulsations
for pulses with a pulsed fraction of greater than 10\% in the period
range from 0.003 to 300 seconds.  Pavlov et al.\ (2002a) do not state
the confidence level to be associated with this upper limit.

\subsubsection{RCW 103 (SNR 332.4$-$0.4)\label{sss:rcw103}}

The source 1E 161348$-$5055 was discovered using the \Einstein\
Observatory by Tuohy and Garmire (1980) and is associated with the SNR
RCW 103 (aka SNR 332.4$-$0.4).  The central point source was
considered to be the prototype of a cooling neutron star (Becker and
Aschenbach 2002) prior to observations with \cha.  Interestingly, the
RCW 103 source was also the first radio-quiet X-ray source found in a
young SNR and, as such, may be considered to be the first detected
CCO.  The source's detailed characteristics (see below) are, however,
somewhat different from the majority of CCOs such as Cas~A and, as has
also been noted by others (e.g. Pavlov, Zavlin, and Sanwal 2002), it
is not clear that all CCOs form a single class of objects.
  
Garmire et al.\ (2000) using \cha\ reported a precise position and
strong evidence for the detection of a significant portion of a
potentially periodic light curve corresponding to a $\approx6$-hr
period based on the 5.5-hr \cha\ observation.  These same authors
reported finding a similar period in archival \ASCA\ data, and
speculated as to the possibility that a binary period had been
detected.  We note that accretion was also listed as one of many
possible explanations for the long term variability observed prior to
the \cha\ observations (see e.g., Gotthelf, Petre, and Vasisht 1999).
 
Binarity may not be the only explanation for the periodicity.  Heyl
and Hernquist (2002) presented the case for a freely-precessing,
short-period neutron star with hot spots as the possible explanation
for the light curve.  Their model predicted that the spin period
should be detectable at minimum flux, but, to date, no detection has
been reported.  Subsequently Becker and Aschenbach (2002) reported the
results of a 5.5-hr \XMM\ observation which showed what appears to be
clear evidence for an eclipse.  If verified, this would the first
clear case of an accreting binary in a SNR.  Further observations of
this interesting system are clearly called for.

Hui and Becker (2005) analyzed these \cha\ data together with the data
from \XMM, fitting the data to two blackbodies and reporting on a
``promising candidate'', albeit statistically weak, pulse period.

\subsubsection{SNR 347.3$-$0.5\label{sss:SNR347.3-0.5}}

Prior to observations of SNR 347.3$-$0.5 with \cha, an X-ray point
source (1WGA J1713.4$-$3949) with no nearby optical counterpart (Slane
et al.\ 1999) had been discovered with \ROSAT\ by Pfeffermann and
Aschenbach (1996).  More interestingly, a radio source (PSR
J1713$-$3945; Crawford et al.\ 2002) with a pulse period of 392 ms was
also detected in the region.  \cha\ observations (Lazendic et
al.\ 2003) played a key role as they were used to establish that the
X-ray source was {\em not} at the same location as the radio pulsar.

Lazendic et al.\ (2003) found that the central X-ray source shares
several characteristics with the CCO in Cas A, including an X-ray
spectrum whose principal component is a blackbody with a temperature
of about 0.4 keV and the absence of pulsations over the frequency
ranges searched (0.01 -- 128 Hz) with upper limits to the pulsed
fraction of better than 25\%.

\subsection{Pulsars with Pulsar Wind Nebulae (PWNe)\label{ss:pwne}}

\cha\ has served more than ably in many cases as a tool to isolate and
resolve the various spatial components that comprise the structure of
pulsar-wind nebulae.  We begin our discussion of the \cha\
contributions to this particular type of study with the remarkable
images that have revealed more observational detail of the complex
interaction between the central pulsar and the surrounding medium than
ever before.  All of the \cha\ PWN images appear to be consistent with
a similar structure --- one that is approximately axially symmetric
and elongated along the symmetry or jet axis.  The most common
explanation identifies the jets with collimated outflows of
relativistic particles along the rotation axis of the underlying
pulsar.  The ubiquitous presence of jets in astrophysical contexts
including these associated with young pulsars and those found with
active galactic nuclei is always worth comment and provides an
over-riding motivation for gathering detailed information to help one
learn about jet formation mechanisms.  Observations with \cha\ have
provided not only spectacular images but also valuable and unique
insights.  One of the most important of these insights is the dynamic
nature of the phenomena that take place in the PWN.  This is
particularly well illustrated in the case of the observations of the
Vela pulsar (\S~\ref{sss:vela}) described by Pavlov et al.\ (2003)
that discovered the variability of the northeast outer jet.

\subsubsection{The Pulsar in the Crab Nebula\label{sss:crab}}

\begin{figure} 
\begin{center}
\resizebox{4in}{!}{\rotatebox{0}{\includegraphics{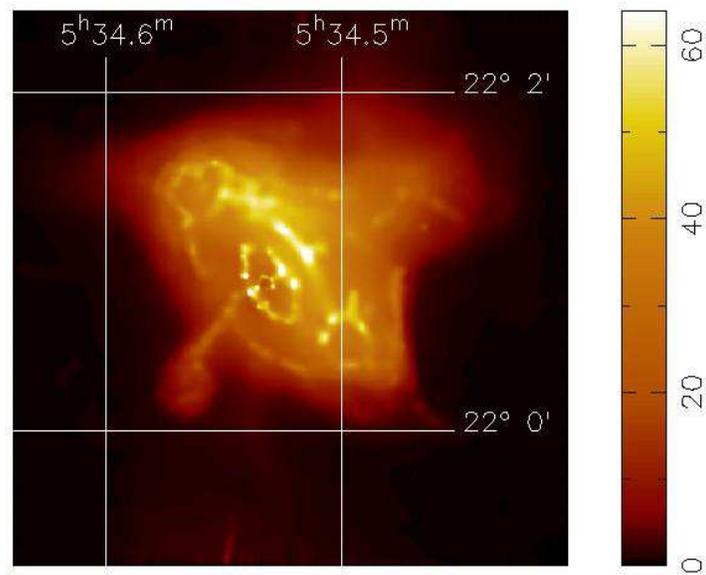}}}
\caption{\label{f:crab_full} \cha\ HETG--ACIS-S3 adaptively smoothed first \cha\ image of the Crab Nebula.
}
\end{center}
\end{figure}

The \cha\ X-Ray Observatory first observed the Crab Nebula and its
pulsar during orbital calibration in 1999.  Weisskopf et al.\ (2000)
published the zeroth-order image obtained with the HETG and read out
by ACIS-S that showed a striking richness of X-ray spatial structures.
Figure~\ref{f:crab_full} shows the original HETG--ACIS-S zeroth-order
image.  A number of features were observed for the first time: an
X-ray inner ring within the X-ray torus; the suggestion of a
hollow-tube structure for the torus; X-ray knots along the inner ring
and (perhaps) along the inward extension of the X-ray jet.  The \cha\
image also clearly resolved the X-ray torus (Aschenbach and Brinkmann
1975) and jet and counterjet which are all features that had been
previously observed (Brinkmann, Aschenbach, \& Langmeier 1985; Hester
et al.~1995; Greiveldinger \& Aschenbach 1999) but never with such
clarity.  On somewhat larger scales, the image showed a sharply
bounded notch (WSW of the Pulsar) into the X-ray nebular emission,
earlier associated with the ``west bay'' of the Nebula (Hester et al.~1995).
Visible-light polarization maps of the Crab Nebula (Schmidt, Angel, \&
Beaver 1979; Hickson \& van den Bergh 1990) demonstrate that the
magnetic field is parallel to the boundary of this notch, thus serving
to exclude the X-ray-emitting relativistic electrons from the west
bay.

The most striking feature of the X-ray image, of course, is the inner elliptical ring, lying between the pulsar and the torus.
The existence of such a ring had been predicted, and corresponds to a shock in the pulsar wind (Rees and Gunn 1974; Kennel and Coroniti 1984).
On the ring reside a few compact knots, and one can be seen in Figure~\ref{f:crab_full} lying SE of the pulsar along the projected inward extension of the jet.
The surface brightness of this knot is too high to be simply explained as the superposition of the ring's and jet's surface brightnesses. 
Ultimately the nature of these knots needs to be probed by means of  high-resolution spectroscopy.

Subsequently, Tennant et al.\ (2001) observed the Crab Nebula and its
pulsar with \cha\ using the Low Energy Transmission Grating
Spectrometer (LETGS).  Time-resolved zeroth-order images were used to
perform a most sensitive search for X-ray emission from the pulser as
a function of pulse phase, including pulse phases that had been
traditionally referred to as ``unpulsed''.  It has been common
practice to denote any minimum in a pulse profile to be representative
of the unpulsed flux; subtract these data from the remainder and then
purport that the difference, represents the ``pulsed'' component.  Of
course this need not be the case.  One can see from the \cha\ image,
for example, that such an approach may include the flux from the inner
ring if not the entire PWN as a whole, depending on the angular
resolution of the instrument.  It is far more satisfactory to use
high-resolution spatial imaging as is possible with \cha\ to isolate
the pulsar from any background that may be present.

Using this approach, Tennant et al.\ (2001) discovered that, as in the
visible (Golden, Shearer, \& Beskin 2000; Peterson et al.\ 1978), the
pulsar emits X-rays at {\em all} pulse phases.  They confirmed prior
observations (Pravdo, Angellini, \& Harding 1997; Massaro et al.\ 2000)
that showed that the power law spectral index varied with pulse phase
and extended the measurements into the pulse minima.  Finally, and,
assuming that all of the flux from the pulsar at pulse minimum is
attributable to thermal emission, the authors used these data to set a
new upper limit to the blackbody temperature.  As a representative
case, they took $\theta_{\infty} = 2.1\times 10^{-16}\, {\rm rad}$~---
for $R_{\infty} = 13\, {\rm km}$ (e.g., $R_{s} = 10\, {\rm km}$ and $M
= 1.4 M_\odot$) at $D = 2\, {\rm kpc}$~--- and $N_{H} = 3\times
10^{21}\, {\rm cm}^{-2}\!$.  With these parameters, the blackbody
temperature that would account for all the flux observed at the pulse
minimum was $T_{\infty} = 2.12\, {\rm MK}= 0.183\, {\rm keV}\!$, which
bounds the actual temperature ($L_{\infty} \approx 2.4 \times
10^{34}\, {\rm erg\ s}^{-1}\!$).  Subsequent \cha-LETG observations
and analyses of the spectrum as a function of pulse phase (Weisskopf
et al.\ 2004) slightly improved this upper limit to $T_{\infty} < (1.76
(2\sigma); 2.01 (3\sigma))$~MK.\footnote{These upper limits appear
weaker than previous \ROSAT-established upper limits set by Becker and
Aschenbach (1995). The \ROSAT\ limits were, however, ``optimistic'' as
discussed in Tennant et al.\ (2001).}

Weisskopf et al.\ (2004) also performed a detailed analysis of the
pulse-averaged spectrum.  They were able to study the interstellar
X-ray extinction due primarily to photoelectric absorption and
secondarily to scattering by dust grains in the direction of the Crab
Nebula.  They confirmed the findings of Willingale et al.\ (2001) that
the line-of-sight to the Crab is under abundant in oxygen.  Using the
abundances and cross sections from Wilms, Allen and McCray (2000) they
found [O/H] = ($3.33\pm 0.25$)$\times 10^{-4}$.  Spectral studies such
as this, where the abundances are allowed to vary, are important as it
is unlikely that standard abundances apply equally well to all lines
of sight, especially those that intersect large quantities of SN
debris (for more on this point see the discussion in Serafimovich et
al.\ 2004 and our comments in \S~\ref{sss:n158a}).

\begin{figure}
\begin{center} 
\resizebox{4in}{!}{\rotatebox{0}{\includegraphics{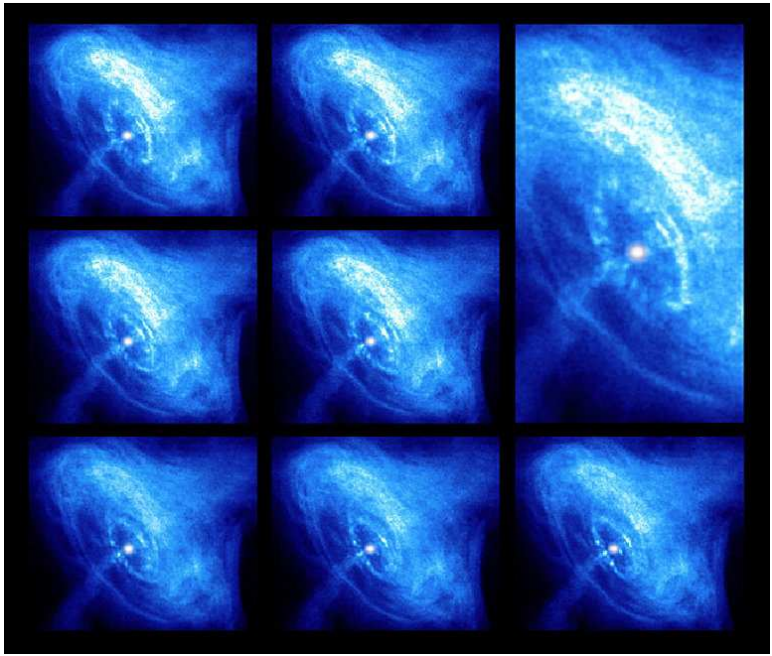}}}
\caption{\label{f:crabmovie} The \cha\ images in this collage were made over a span of several months and are 1.6' x 1.6' and are time-ordered left to right and top to bottom except for the larger close-up. Image courtesy NASA/CXC/ASU/ J. Hester et al. (2002).
}
\end{center}
\end{figure}

In 2002, Hester et al.\ (2002) completed one phase of a set of
coordinated observations of the Crab's PWN using \cha\ (ACIS-S in
sub-array mode) and the Hubble Space telescope.  These spectacular
observations revealed numerous dynamical features including wisps
moving outward from the inner equatorial X-ray ring at about 0.5 c.
The eight \cha\ observations are shown in Figure~\ref{f:crabmovie}.

Finally, \cha\ (and \XMM, see Willingale et al.\ 2001) has been used to
study spectral variations as a function of position in the nebula.
Weisskopf et al.\ (2000) first presented the variation of a hardness
ratio (the ratio of flux in two energy bands) as a function of
position as seen with \cha\ using 5" x 5" pixels.  Mori et al.\ (2004)
followed this work with studies of the variation of the power law
spectral index as a function of position using 2.5" x 2.5" pixels and
the same data discussed in Hester et al.\ (2002).  Despite the
particular mode (subframe) used to obtain these data, the effects of
pulse pileup plagued the data analysis and required the application of
corrections.  These corrections, at best, did not adequately correct
for pileup when bright spatial structure was present within an
analysis pixel --- dealing with that particular situation was noted by
the authors to be beyond the scope of the paper.  One hopes this
problem will be be addressed by some enterprising expert in pulse
pileup in the future, since it is at the smallest spatial scales that
the \cha\ observations are most unique.  Performing spectroscopy
(perhaps even time-resolved spectroscopy) of the bright, compact
features is necessary in order to understand their origin.
 
\subsubsection{The Vela Pulsar and its remarkable external jet\label{sss:vela}}

\begin{figure} 
\begin{center}
\resizebox{4in}{!}{\rotatebox{0}{\includegraphics{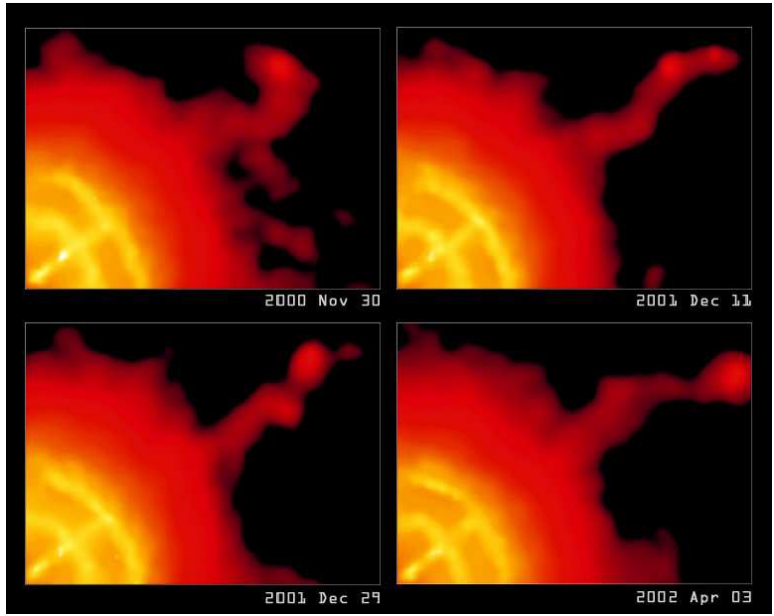}}}
\caption{\label{f:vela} \cha\ ACIS-S3 images in a montage showing the variability in both intensity and position of the jet associated with the Vela pulsar. The pulsar is located at the lower left-hand corner of each image. These four images are part of a series of 13 images made over a period of two and a half years. Each image is 1.6' x 1.2'. Image courtesy NASA/CXC/PSU/G. Pavlov et al. (2003).
}
\end{center}
\end{figure}

\cha\ observations of the 89-ms period Vela pulsar (Helfand, Gotthelf,
\& Halpern 2001; Pavlov et al.\ 2003) and its surroundings have been
most revealing.  In addition to showing the complex and time variable
spatial structure of the region immediately surrounding the pulsar
itself --- a structure that includes two sets of arcs, a jet in the
direction of the pulsar's proper motion and a counterjet --- the \cha\
images taken by Pavlov et al.\ (2003) also discovered that the
continuation of the jet that extends to the NW is time variable in
both intensity and position on scales of days to weeks as illustrated
in Figure~\ref{f:vela}.  The brightness variations are different along
the extent of the jet, and Pavlov et al.\ (2003) inferred flow
velocities of 0.3--0.7c.  Finally, the apparent width of the 
outer jet appears to be constant, despite large variations in
appearance, indicating confinement.  The analogy to a fire hose being
held at its base appears very appropriate.

The \cha\ Observatory should be used to both to search for, and to study, such behavior in all cases where it is clear that jets from neutron stars are present. 

\subsubsection{PSR1509$-$58\label{sss:psr1509}}

\begin{figure} 
\begin{center}
\resizebox{4in}{!}{\rotatebox{0}{\includegraphics{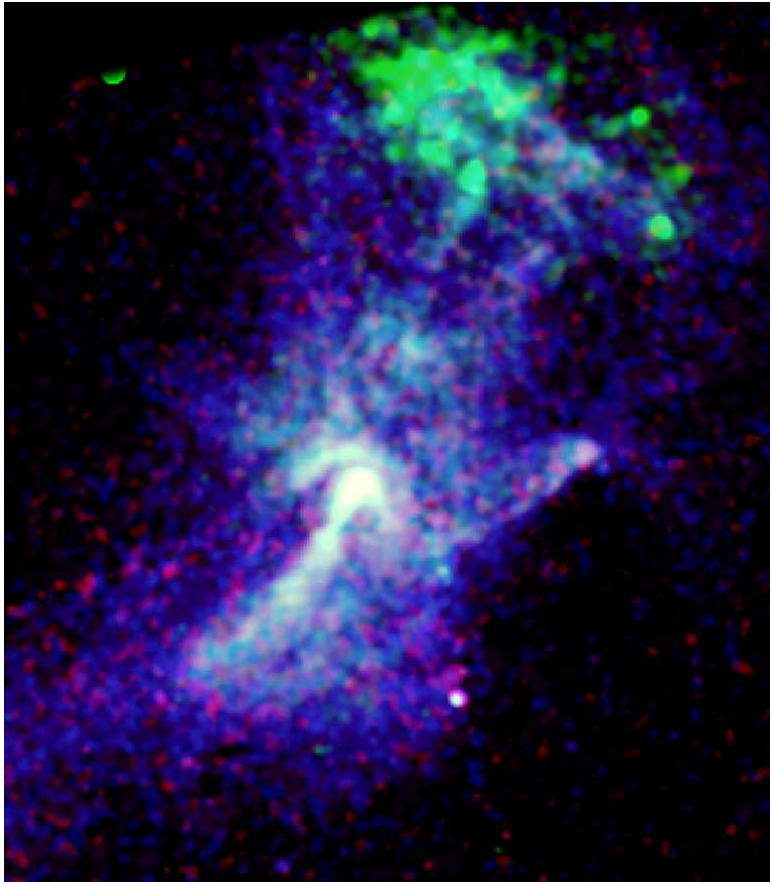}}}
\caption{\label{f:B1509-58} \cha\ ACIS-I image of pulsar B1509$-$58 in SNR SNR 230.4$-$1.2. The image is 10' by 14'. The pulsar is the bright white source at the center of the nebula. Courtesy NASA/MIT Gaensler et al. (2003).
}
\end{center}
\end{figure}

The \cha\ image of the young PWN powered by B1509$-$58 in SNR
230.4$-$1.2 (Gaensler et al.\ 2002) is shown in
Figure~\ref{f:B1509-58}.  The pulsar is the bright source at the
center of the nebula.  A thin jet can be seen in the image to extend
to the southeast.  Just above the pulsar there is a small arc of X-ray
emission, which seems to mark the location of the shock wave produced
by the particles flowing away from the pulsar's equator.  The cloud
near the top of the image may to be due to high temperature gas.  This
gas, possibly a remnant of the explosion associated with the creation
of the pulsar, may have been heated by collisions with high-energy
particles produced by the pulsar.  See Yatsu et al.\ (2005) for a
discussion of the interaction of the pulsar's jet with this material.

\subsubsection{SNR 292.2$-$0.5\label{sss:SNR292.2}}

SNR 292.2$-$0.5 contains the 407-ms radio pulsar J1119$-$6127 whose
discovery (Camilo et al.\ 2000) led to the radio detection of the SNR
by Crawford et al.\ (2001).  The first of two \cha\ observations with
ACIS-S3 (Gonzalez and Safi-Harb 2003, 2005) provided unambiguous
detection of the X-ray counterpart to the radio pulsar and strong
evidence for the detection of a faint PWN.  Gonzalez and Safi-Harb
(2003, 2005) found that the combined X-ray emission from the pulsar
and its associated nebula is described by an absorbed power law model
with a photon index 2.2 (+0.6, $-$0.3) and an unabsorbed X-ray
luminosity (0.5--10.0 keV) of $5.5 (+10, $-$3.3) \times 10^{32}$ ergs
s$^{-1}$ assuming a 6 kpc distance.  An interesting attribute is the
source's derived inefficiency in converting rotational energy into
X-rays using the standard assumptions.  Deeper observations are need
to better establish the detailed attributes of the PWN, such as
spectral variations as a function of position.

\subsubsection{SNR 54.1+0.3\label{sss:SNR54.1}}

\begin{figure} 
\begin{center}
\resizebox{4in}{!}{\rotatebox{0}{\includegraphics{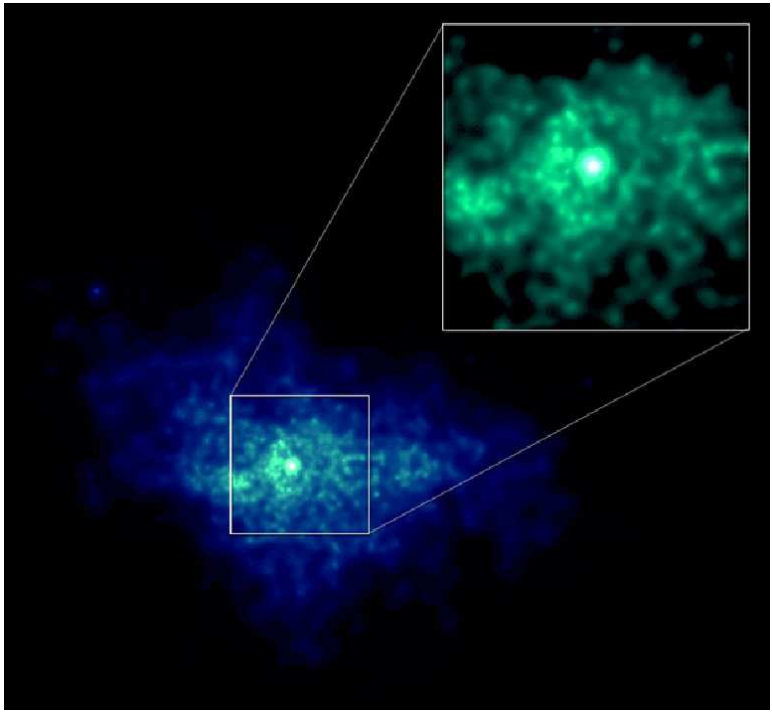}}}
\caption{\label{f:SNR54.1+0.3} \cha\ ACIS-S3 image of SNR 54.1+0.3. The large image is 2.7' x 2'. Courtesy NASA/CXC/U.Mass/F. Lu et al.
}
\end{center}
\end{figure}

Lu et al.\ (2002) observed SNR 54.1+0.3 using ACIS-S3.  The \cha\
image is shown in Figure~\ref{f:SNR54.1+0.3} and one sees a central
bright pointlike source, a surrounding ring, jet-like elongations, and
low surface brightness diffuse emission.  All of these features
emphasize the similarity to the Crab's PWN.  Lu et al.\ (2002)
determined that the spectra of these components are all well described
by power-law models (as with the Crab PWN); the spectral index
steepens (softens) with increasing distance from the point source.
The similarity of SNR 54.1+0.3 to the Crab nebula and its pulsar was
further strengthened by the subsequent discovery (Camilo et al.\
2002a) of a 136-ms radio pulsar at the location of the \cha\ source.
(These authors then also detected the pulsations in archival X-ray
observations performed with the \ASCA\ satellite.)  The radio pulsar,
PSR J1930+1852, is very weak with a period-averaged flux density at
1180 MHz of 60 $\mu$-Jy.  For a distance of 5 kpc, the corresponding
luminosity is among the lowest for known young pulsars.

\subsubsection{SNR 39.2$-$0.3 (3C 396)\label{sss:SNR39.2-0.3}}

\begin{figure}
\begin{center} 
\resizebox{4in}{!}{\rotatebox{0}{\includegraphics{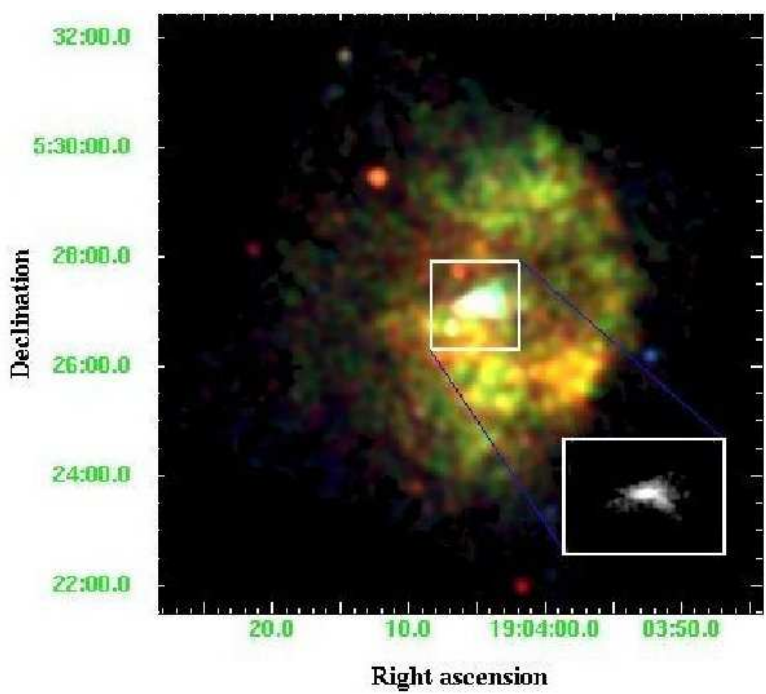}}}
\caption{\label{f:3C396} \cha\ ACIS-S3 image of 3C 396.  Courtesy M. J. Rutkowski and J. Keohane.
}
\end{center}
\end{figure}

Olbert et al.\ (2003) observed SNR 39.2$-$0.3 (3C 396) using ACIS-S3.
The \cha\ image (Figure~\ref{f:3C396}) resolved an extended
($55^{\prime\prime} \times 20^{\prime\prime}$) X-ray nebula with a
nonthermal energy spectrum ($\Gamma = 1.5 \pm 0.3$ at 90\%-confidence)
and detected what the authors refer to as a ``pointlike'' source at
the center of the nebulosity.  There is also diffuse radio emission in
the same region.  This discovery provided convincing evidence for the
presence of a PWN, surely harboring an X-ray pulsar with a
to-be-detected pulse period.  The \cha\ observations confirmed the
conclusions as to the nature of this source --- the existence of both
thermal and non-thermal components and the possible presence of a
rotating NS powering a synchrotron nebula --- reached by Harrus and
Slane (1999) based on observations with \ASCA.  The pulse period, if
observable in our line of sight, is yet to be detected and Olbert et
al.\ (2003) do not appear to have set upper limits over the admittedly
long periods accessible using ACIS in its normal mode.  A recent
search (Zavlin 2005) of these data and covering the period range from
10 to 10$^4$ s uncovered no evidence for pulsations for pulses with a
sinusoidal amplitude of greater than 34\% (95\%-confidence).

\subsubsection{SNR 293.8+0.6\label{sss:SNR293.8+0.6}}

\begin{figure} 
\begin{center}
\resizebox{5in}{!}{\rotatebox{0}{\includegraphics{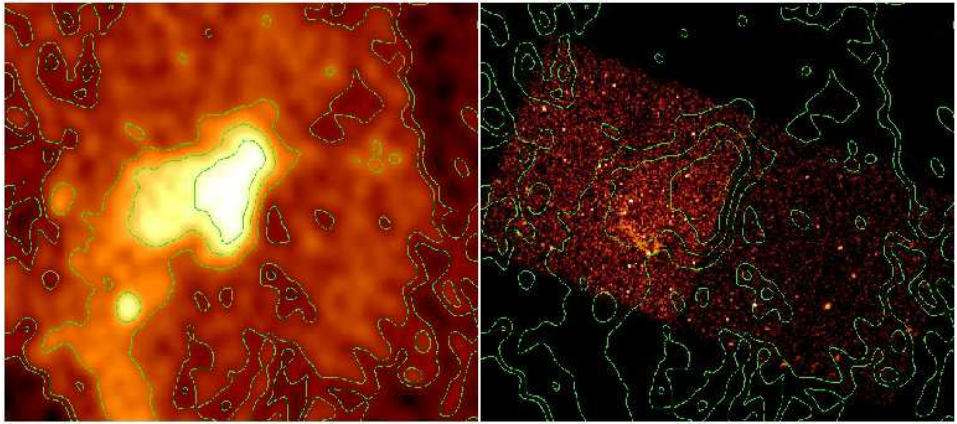}}}
\caption{\label{f:SNR293.8+0.6} Radio (left) and \cha\ (right) with radio contours superimposed of SNR 293.8+0.6. Image courtesy S. Patel.
}
\end{center}
\end{figure}

SNR 293.8+0.6 was observed with \cha\ ACIS-S3 for 40 ksec by Olbert,
Keohane, and Gotthelf (2003).  There is no published reference to this
observation other than the abstract referred to here.  The abstract
mentions the presence of a ``soft'' point source near the center of
the remnant and the absence of a bright synchrotron nebula.  The
authors note that these results seem to be in contrast to the presence
of a PWN that one might expect from the radio image.  A quick glance
at the \cha\ and radio images together (shown in
Figure~\ref{f:SNR293.8+0.6}) however, shows numerous X-ray sources in
the region containing the SNR, any of which might be candidates for an
associated compact object.  The X-ray image in
Figure~\ref{f:SNR293.8+0.6} is rich in structure and clearly more work
needs to be done before one adds SNR 293.8+0.6 to the list of SNR with
established and identified compact objects.

\subsubsection{N158A (SNR B0540$-$69 in the LMC)\label{sss:n158a}}

\begin{figure} 
\begin{center}
\resizebox{5in}{!}{\rotatebox{-90}{\includegraphics{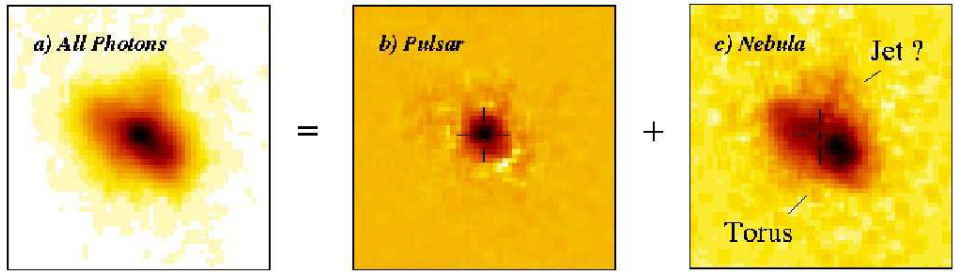}}}
\caption{\label{f:B0540-69} \cha\ HRC-I images of the pulsar wind nebula of B0540$-$69 in the LMC. These images are approximately 12.5" x 12.5". Images courtesy E. Gotthelf.
}
\end{center}
\end{figure}

The 50-ms pulsar PSR B0540$-$69 is very ``Crab-like'' possessing a
similar pulse period, spin-down age, and spin-down power.  The pulsar
was discovered by Seward, Harnden, and Helfand (1984) using the
\Einstein\ Observatory.  A \cha\ HRC observation, performed by
Gotthelf and Wang (2000), was motivated by a desire to search for an
X-ray plerion, expected based on the similarity to the Crab and
previous observations in the optical, and radio (see e.g. the
introduction in Gotthelf and Wang (2000) and references therein).
Using the HRC allowed for precision timing and subsequent detection of
the pulsed emission from the central, barely-resolved, extended
emission.  Defining the data at pulse minimum, which lasted about 0.5
in pulse phase, as ``off-pulse'' these authors separated the central
image into that of the ``pulsed'' component and that for the
``off-pulse'' component.  The former appeared point-like whereas the
latter appeared extended as shown in Figure~\ref{f:B0540-69} revealing
the presence of a PWN and even possibly a jet.

Kaaret et al.\ (2001), in addition to reanalyzing the HRC data and
applying an improved aspect solution and improved HRC processing to
remove spatial artifacts, also performed ACIS-I measurements in
continuous clocking mode.  As noted previously, this mode allows for
higher time resolution albeit at the price of one spatial dimension in
the image.  Thus, the data available for spectral analysis had to
contend with a significant contribution, not so much from the PWN but
from the larger SNR.  The spectrum of the pulsed component was found
to be consistent with a power law with a photon index of 1.83 ($\pm
0.13$) fixing N$_{H}$ at $4.6 \times 10^{21}$ cm$^{-2}$.  Spectral
analyses of the nebula by Kaaret et al.\ (2001), again fixing N$_{H}$,
showed it to be softer than the pulsar with spectral indices varying
from 1.85 to 2.26 although it is not totally clear that the variations
are that statistically significant.  The variation is in contrast to
the lack of a spectral variation over the PWN measured in the visible
(Serafimovich et al.\ 2004).

Serafimovich et al.\ (2004) also made important contributions to the
study of this system analyzing both new (VLT) and archival (HST)
observations.  Moreover, these authors raised the important point,
which to our knowledge has been neglected when analyzing spectra
(except in the case of the Crab pulsar --- as discussed in
\S~\ref{sss:crab}) that the use of standard abundances may not be (is
not) justified, perhaps even more so for an extragalactic source.  Of
general interest (although not necessarily directly applicable to
B0540$-$69 as they noted) is their discussion of the potential impact of
the absorption produced by the SN ejecta on the X-ray spectrum which
can enhance the X-ray absorption at energies above the oxygen K-edge
depending on the state of evolution of the SNR --- the less evolved,
the more possible that there is a significant contribution to the
absorption from the ejecta.  {\em We strongly urge readers interested
in understanding the spectra of X-ray sources in SNR to carefully
examine the discussion in \S~2.7.1 of Serafimovich et al.\ (2004).}

Serafimovich et al.\ (2004) did note that the pulsar could well have a
non-power-law (thermal?) spectrum at energies below 1 to 2 keV,
although this conclusion must be tempered by the knowledge that the
understanding of the ACIS response function decreases with decreasing
energy.  Serafimovich et al.\ (2004) concluded that a reanalysis of the
X-ray spectrum of B0540$-$69 is called for, but noted that this was
beyond the scope of their paper.  The reanalysis should be done, and
similar considerations such as allowing for absorbing effects of
ejecta, allowing abundances to vary, etc. need to be applied in a
systematic way to all spectral analyses performed in the \cha\ era
where not prevented by poor statistics.  Indeed, it might be argued
that spectral observations should not be performed unless there are
sufficient statistics to pursue such studies --- simply being able to
differentiate between a power law and say a blackbody spectrum, all
else being equal, may no longer be sufficient justification for
establishing the length of such observations.

\subsubsection{N157B (NGC 2060; SNR 0538$-$691) \label{sss:n157b}}

\begin{figure} 
\begin{center}
\resizebox{4in}{!}{\rotatebox{90}{\includegraphics{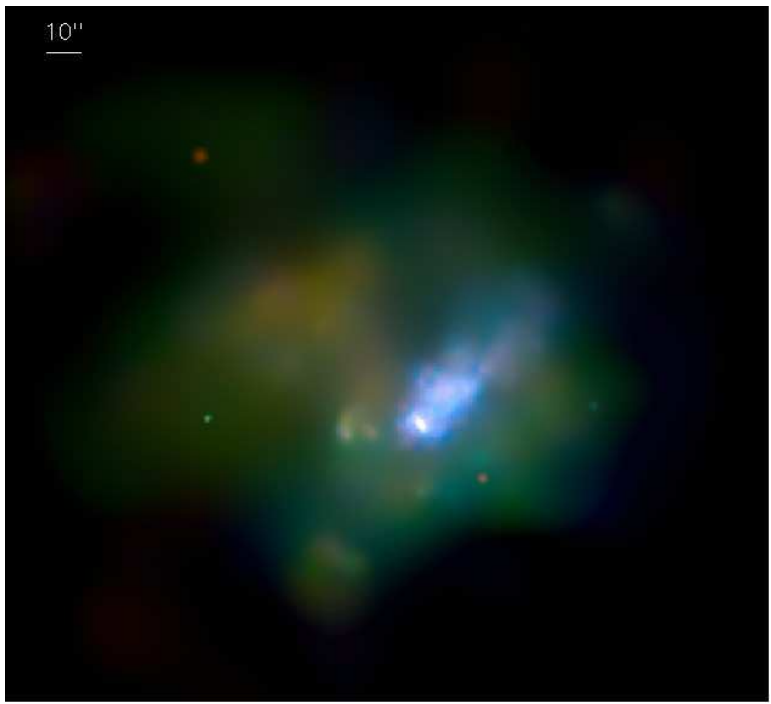}}}
\caption{\label{f:n157b} \cha\ image of SNR N157B. The image combines HRC-S and -I observations and is smoothed to bring out the extended structure near the pulsar. Courtesy Q.D.Wang.
}
\end{center}
\end{figure}

The supernova remnant N157B (aka NGC 2060, SNR 0538$-$69.1, and 20
Doradus B) contains the RXTE-discovered (Marshall et al.\ 1998)
16-ms-period X-ray pulsar, PSR J0537$-$6910 and is located in the LMC.
\cha\ observations using the HRC-I and HRC-S by Wang et al.\ (2001)
served a number of purposes including obtaining a precise location for
the pulsar.  In addition they were able to spatially resolve the
pulsar, a surrounding compact yet elongated (~$0.6\times 1.7$ pc)
nebula, and an even larger-scale feature of diffuse emission trailing
from the pulsar and oriented nearly perpendicular to the major axis of
the nebula.  These features, shown in Figure~\ref{f:n157b}, indicate
interesting interactions between the pulsar-powered nebula as it is
moving through the surrounding medium.  A subsequent ACIS-S
observation (ObsID 2783) seems never to have been formally analyzed.

\subsubsection{B0453$-$685\label{ss:B0453}}

\begin{figure} 
\begin{center}
\resizebox{4in}{!}{\rotatebox{0}{\includegraphics{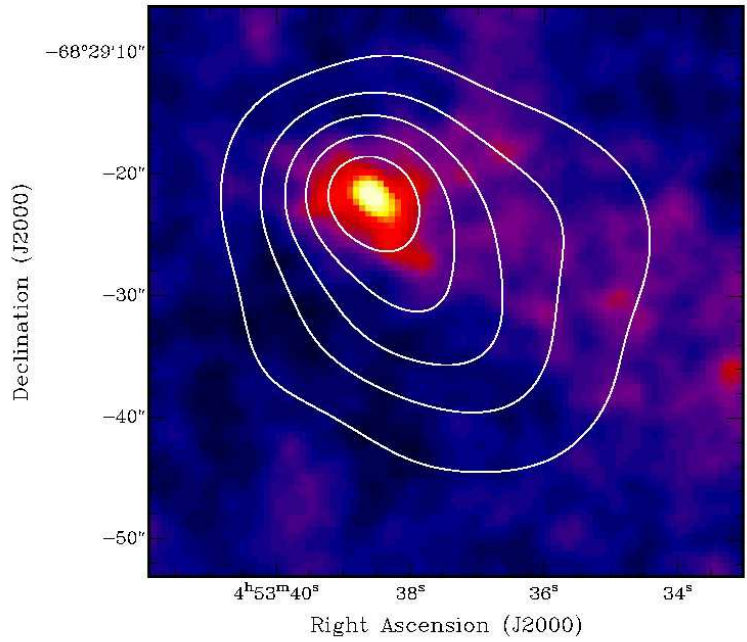}}}
\caption{\label{f:B0453} \cha\ ACIS-S3 image of the pulsar wind nebula and corresponding radio contours of B0453$-$685 in the LMC. Image courtesy B. Gaensler.
}
\end{center}
\end{figure}

Gaensler et al.\ (2003) performed both radio and \cha\ observations of
the supernova remnant B0453$-$685 in the Large Magellanic Cloud (LMC) and
discovered a new PWN.  Gaensler et al.\ (2003) detected a strongly
linearly polarized (8\% and 6\% at 2.4 GHz and 1.4 GHz respectively)
and elongated ($20\arcsec \times 30\arcsec$) central radio core,
similar in morphology to the X-ray core ($14\arcsec \times 7\arcsec$)
seen as part of a 40 ksec ACIS-S3 observation (Figure~\ref{f:B0453}).
They also found that the X-ray spectrum of the central core is fit by
a power-law with number index of $-1.9 \pm 0.4$.  Unfortunately no
mention is made as to the quality of the fit.  This is important as
the statistical uncertainties (measured by extremes on the $\chi^{2} +
\epsilon$ contours) get {\em underestimated} if the fit is poor.
Based on the quoted uncertainties and the limited number of counts, we
suspect that the fit to the power-law spectrum may not have been
compelling in comparison to other models for the continuum.  However,
the radio and X-ray images, and their similarity leave little doubt
that a PWN has been detected.  The search for the underlying pulsar
proved unrewarding, albeit not surprisingly so, as these authors found
a scant 58-count upper limit to the contribution of a putative point
source to the central image.  The authors noted that the corresponding
limits to the X-ray luminosity (0.5--10.0 keV) of $< 6 D^{2} \times
10^{33}$ ergs s$^{-1}$ (assuming $N_H = 1.3 \times 10^{21}$ cm$^{-2}$
and a power law index $\Gamma = 1.5$) were hardly restrictive.  It is
worth emphasizing that Gaensler et al.\ (2003) propose an interesting
approach to determining the properties of the system in contrast to
the usual one where the equations for a pulsar's age, spin-down
luminosity, and surface magnetic field are written down in the dipole
approximation and one assumes a braking index of 3, an initial period
of 0, and a luminosity that is a carefully selected fraction of the
spin-down luminosity in order to determine the pulse period, period
derivative, and the neutron star's surface magnetic field.  We won't
repeat their discussion here, but refer interested readers to \S~3.1
of their paper.

Deeper observations of this system, which would allow one to perform
spatially resolved spectroscopy in addition to facilitating sensitive
searches for pulsations, are clearly called for.

\subsubsection{CTA 1 (SNR 119.5+10.2)\label{sss:cta1}}

CTA 1 (aka SNR 119.5+10.2) is a radio shell SNR with emission from the
center that appeared harder and brighter than the emission from the
limb as seen with \ROSAT\ (Seward, Schmidt, \& Slane 1995).  Of
special interest is the fact that the position of 3EG J0010+7309, one
of the brighter of the EGRET unidentified sources (Hartman et al.\
1999), lies inside the boundary of CTA 1.

\begin{figure} 
\begin{center}
\resizebox{4in}{!}{\rotatebox{0}{\includegraphics{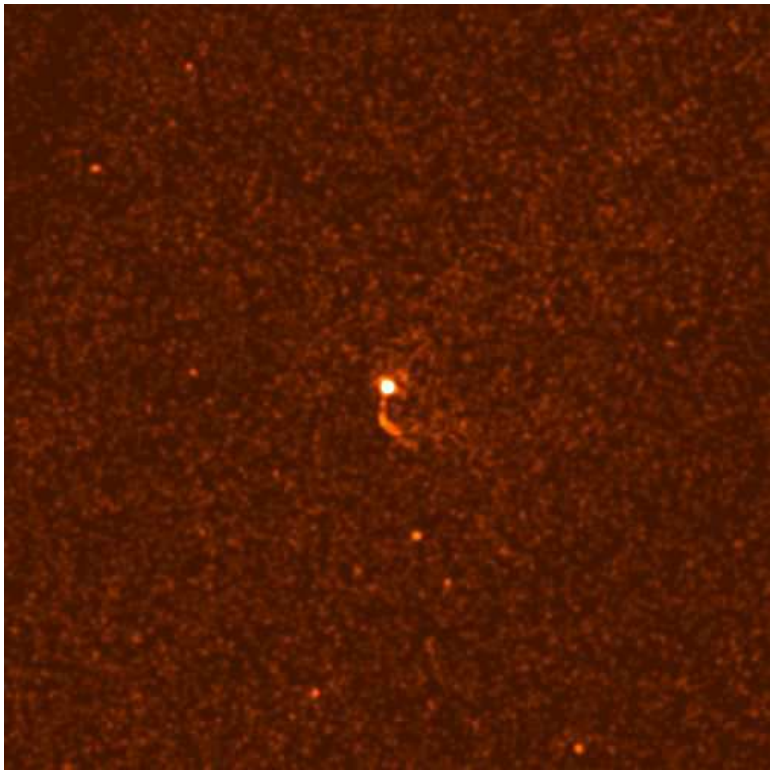}}}
\caption{\label{f:cta1} \cha\ ACIS-S3 image of CTA 1. The image is 2' x 2'. Courtesy J. Halpern.
}
\end{center}
\end{figure}

Halpern et al.\ (2004) used \cha\ ACIS-S3 to image the central portion
of the remnant.  The \cha\ image (Figure~\ref{f:cta1}) reveals a point
source, a compact nebula, and a (curved) ``jet''.  These
characteristics, together with upper limits to the optical flux at the
\cha\ location of the point source which gives an X-ray-to-optical
flux ratio in excess of 100, clearly establish the central object as
as a rotation-powered pulsar, albeit pulsations are yet to be
detected.

Slane et al.\ (2004a) using \XMM\ set a rather unrestrictive upper
limit of 61\% pulsed fraction for periods between 1.2 ms and tens of
ksec.  Halpern et al.\ (2004) established a restrictive upper limit to
a possible radio counterpart at both 1425 and 820 MHz (implying less
than 0.02 mJy kpc$^2$ at 1400 MHz).  Halpern et al.\ (2004) also found
that the X-ray spectrum of the point source is best fit by a power law
plus blackbody model with $\Gamma = 1.6$, kT$_\infty = 0.13$ and
R$_\infty = 0.4$ km.  We wish to emphasize that the number of counts
from the point source (and the PWN and the jet) were quite limited
being only 187 (136 and 45) implying large uncertainties in the
derived parameters, and indicting that a deeper \cha\ observation is
called for since the angular scale of the observed features is far to
small to be adequately resolved with \XMM.

Finally, Halpern et al.\ (2004) derived an upper limit to the surface
temperature of the underlying neutron star.  The upper limit was
perhaps conservative in that it only made use of the data from the
lowest energy bin rather than examining the full spectrum.  On the
other hand, the column was fixed and allowed variations of the column
might possibly increase the upper limit of $6.6 \times 10^5$ K.

In general there is a real need for a systematic presentation
(e.g. was $N_{\rm H}$ fixed or not, what is the confidence level
associated with the limit? etc.) of upper limits to neutron star
surface temperatures when confronting theory.

\subsubsection{SNR 69.0+2.7 (CTB 80)\label{sss:SNR69.0+2.7}}

\begin{figure} 
\begin{center}
\resizebox{4in}{!}{\rotatebox{-90}{\includegraphics{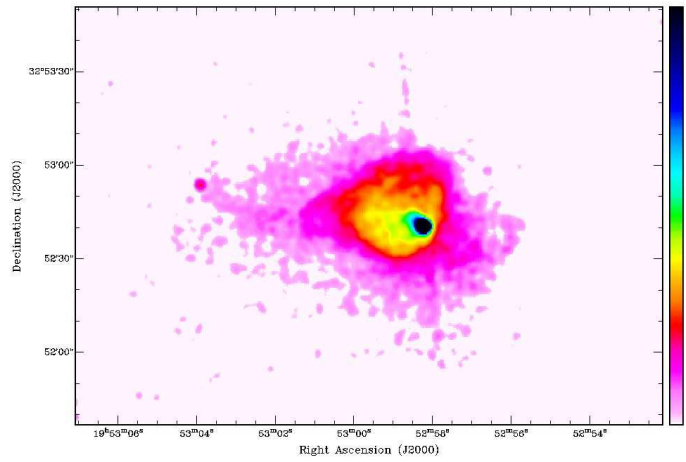}}}
\caption{\label{f:ctb80} \cha\ ACIS-S3 image of CTB 80. Courtesy D-S Moon.
}
\end{center}
\end{figure}

Moon et al.\ (2004) observed the SNR CTB 80 (aka SNR 69.0+2.7) with
\cha\ ACIS-S3 (see Figure~\ref{f:ctb80}) as part of a multi-wavelength
study.  The remnant contains the 40-ms radio pulsar PSR B1951+32
discovered by Kulkarni et al.\ (1988).  The \cha\ observations
clarified the morphology of the X-ray emission and showed what appears
to be a cometary PWN elongated (about 40") along the direction of the
pulsar's proper motion and seemingly confined by a bow shock produced
by that proper motion and thus confirming prior speculations (see
e.g. Moon et al.\ 2004 and references therein).
  
More recently Li, Lu, and Li (2005) have further analyzed these same
data with emphasis on providing spatially-resolved spectra.  Li, Lu
and Li (2005) find that a power-law plus blackbody model fits the
spectrum of the pulsar better than a pure power-law model.  In this
case the blackbody comprises about 10\% of the total flux (with large
uncertainties) and a temperature of 0.13 keV.  The accompanying power
law index was about 1.6.  The corresponding blackbody radius was
small, about 2 km at the assumed distance of 2 kpc, perhaps implying
emission from hot spots.  These authors also fit the data to a
blackbody plus power law, while fixing the blackbody radius to a value
more appropriate to the entire star in an effort to set an upper limit
to the surface thermal emission.  This approach led to a temperature
upper limit ($3\sigma$) of $7.8 \times 10^{5}$ K, much below the
predictions of standard neutron star cooling models (see Figures 8 and
9 of Li, Lu, and Li 2005) as with 3C58 (\S~\ref{sss:3c58}).  The time
resolution of the data did not permit an analysis for 40-ms
pulsations.

\subsubsection{SNR 359.23$-$0.82 (The Mouse)\label{sss:mouse}}

\begin{figure} 
\begin{center}
\resizebox{4in}{!}{\rotatebox{-90}{\includegraphics{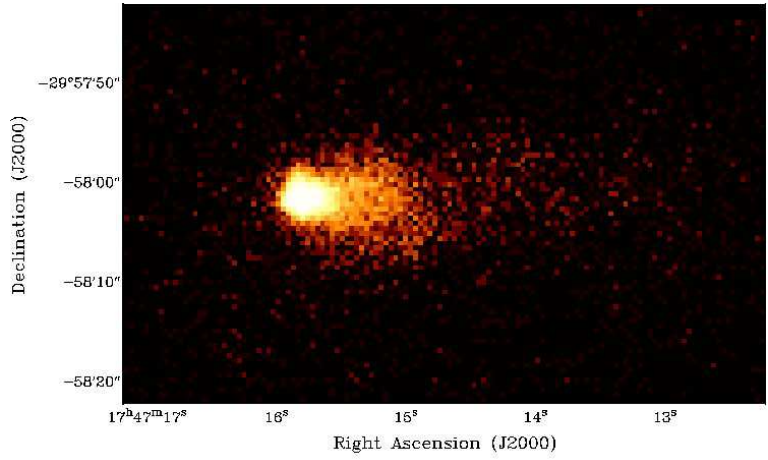}}}
\caption{\label{f:mouse} \cha\ ACIS-S3 image of the "mouse". Courtesy B. Gaensler.
}
\end{center}
\end{figure}

Gaensler et al.\ (2004) observed the unusually shaped radio source SNR
359.23$-$0.82, sometimes referred to as the ``mouse'', using ACIS-S3.
For a history of observations of this object in all wavelength bands
see the introduction to Gaensler et al.\ (2004) and references therein.
Observations with \ROSAT\ by Predehl and Kulkarni (1995) had already
detected X-ray emission and they correctly proposed that this source
was a bow shock PWN.  The \cha\ observations confirmed this conclusion
and, as shown in Figure~\ref{f:mouse}, provided spectacular details
resolving numerous components referred to as the ``halo, head, tongue
and tail'' by Gaensler et al.\ (2004).  We note that Gaensler et
al.\ (2004) goes well beyond the presentation of new observational
results and they combine theory and hydrodynamic simulations of bow
shocks to unravel the implications of the data.  The paper is must
reading for those interested in the interaction of a PWN with the
environment, especially when the pulsar has a moderately large
velocity.

\subsubsection{Geminga\label{sss:geminga}}

\begin{figure} 
\begin{center}
\resizebox{4in}{!}{\rotatebox{0}{\includegraphics{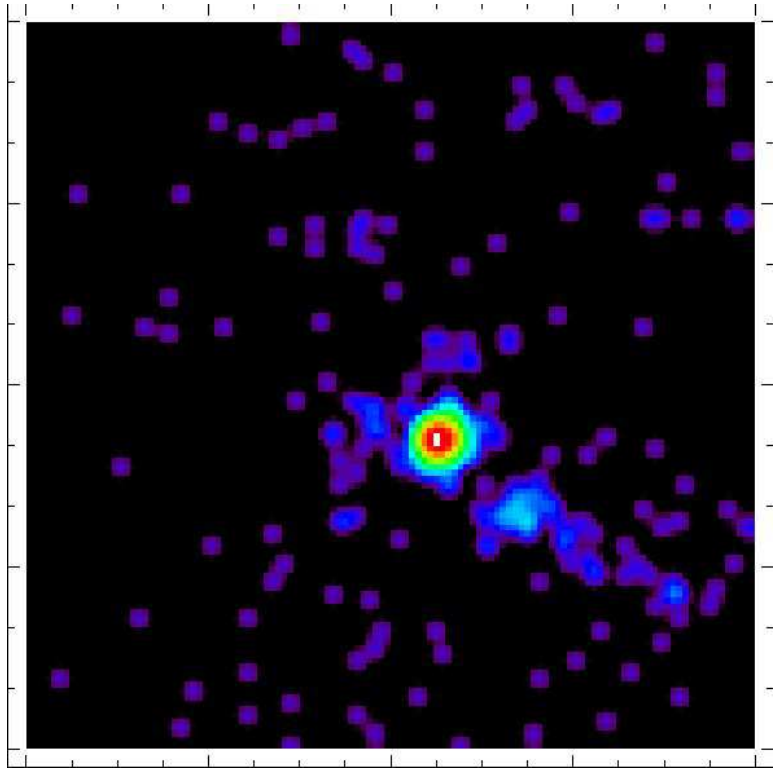}}}
\vspace{0.15in}
\caption{\label{f:geminga} \cha\ ACIS-S3 image of Geminga. Courtesy D. Sanwal \& V. Zavlin.
}
\end{center}
\end{figure}

Geminga was observed using ACIS-S3 by Sanwal, Pavlov and Zavlin
(2004).  The resulting image is shown in Figure~\ref{f:geminga} where
one sees what Sanwal, Pavlov, and Zavlin (2004; 2005) term a ``wake'',
about 10$^{\prime\prime}$ -- 15$^{\prime\prime}$ in projected length
to the southwest of the pulsar.  If one looks carefully, one can also
see extended emission a few arc seconds away from the pulsar in the
opposite direction, perhaps the head of a bow-shock.
   
\subsection{Finding the pulsars\label{ss:findingpulsars}}

\subsubsection{SNR 296.5+10.0 and 1E 1207.4$-$5209\label{ss:SNR296.5+10}}

Observations with \cha\ have contributed at least two significant new
insights into the source 1E 1207.4$-$5209 initially discovered with
the \Einstein\ Observatory (Helfand \& Becker 1984) and located 6'
from the center of SNR PKS 1209$-$51/52 (aka SNR 296.5+10.0).  The
source was first observed with \cha\ by Zavlin et al.\ (2000) using
ACIS-S3.  These observers used ACIS in continuous clocking mode which
allows time resolution of 2.85 ms at the price of one dimension of
spatial information and discovered a 424-ms period.  The detection of
the period, of course, provided compelling evidence that the source is
a neutron star.  Since the source appears to be radio-quiet
(Mereghetti, Bignami, \& Caraveo 1996; Kaspi et al.\ 1996), it may be
either an active pulsar beamed out of our line of sight or a truly
radio-quiet neutron star, where the X-ray pulsations are caused
perhaps by hot spots rotating in and out of our line of sight.

Subsequent to the detection of pulsations, Sanwal et al.\ (2002)
analyzed two ACIS-S3 continuous-clocking-mode observations including
the data used to initially detect the pulse period.  In addition to
establishing a preliminary estimate for the period derivative, these
authors also found two significant absorption features centered at 0.7
and 1.4 keV with equivalent widths of about 0.1 keV.  Sanwal et
al.\ (2002) discussed several possible interpretations for the
absorption including cyclotron resonances and atomic features.  They
presented arguments favoring atomic transitions of once-ionized helium
in the atmosphere of the neutron star assumed to be very strongly ($
\approx 10^{14}$G) magnetized.  The exact cause of the \cha-discovered
features is not without different interpretations.  For example Hailey
and Mori (2002) argued that the absorption features were associated
with He-like oxygen or neon in a field of $\approx 10^{12}$G.  More
recent observations with \XMM\ (e.g., Mereghetti et al.\ 2002; Bignami
et al.\ 2003; De Luca et al.\ 2004) not only confirmed the \cha-detected
absorption features at 0.7 and 1.4 keV, but also seemed to have
uncovered an additional feature at 2.1 and evidence for a fourth 2.8
keV.  Taking all these latter data into account supports an
explanation involving the fundamental and two, possibly three,
harmonics of the electron cyclotron absorption in a field of order
$10^{11}$G.  However, the two additional spectral features in the
\XMM\ data have not been unambiguously accepted.  Mori, Chonko,
and Hailey (2005) have cast severe doubt as to the reality of the spectral features at 2.1 and 2.8 keV.  The arguments given
seem compelling and it is thus unfortunate that the \cha\ response is
insufficient to weigh in on this question without expending
significant amounts of observing time.

Zavlin et al.\ (2004) have continued to observe this target using both
\cha\ and \XMM.  They have detected significant variations in
the spin period, which they interpreted in light of three hypotheses:
a glitching pulsar; variations in an accretion rate from a fallback
disc; and variations in accretion produced by being in a wide binary.

Thus the sequence of \cha\ observations have provided important
discoveries, especially the detection of the pulse period and firm
detection of two absorption features.  An important and unanswered
question is what are the limits as to the presence of such spectral
features for the other NSs in SNR.  A systematic comparison, if not
already in progress, should be performed.

Finally we note that 1E 1207.4$-$5209 is a source that, in some
critical respects, is similar to Cas~A in that it is in a SNR, is
radio-quiet, and has a low-energy spectrum that may be fit by a
blackbody with a temperature falling in the range from 0.2--0.6 keV,
however it pulses.  Thus, on the one hand, the source therefore can be
used to give us confidence that all CCOs will ultimately be found to
pulse.  On the other, this source's \cha-revealed characteristics may
be used to separate it from the CCO-group of objects.

\subsubsection{SNR 292.0+1.8\label{sss:SNR292.0}}

\begin{figure} 
\begin{center}
\resizebox{4in}{!}{\rotatebox{0}{\includegraphics{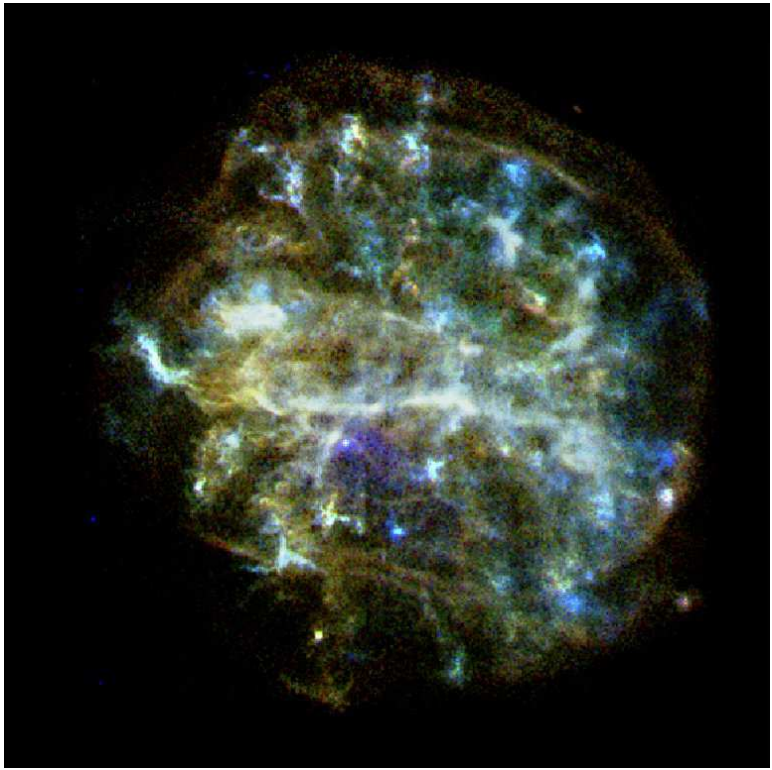}}}
\caption{\label{f:SNR292.0+1.8} \cha\ ACIS-S3 image of SNR 292.0+1.8. The image is 9' x 9'. The point source identified with the 135-ms PSR J1124$-$5916 is just southeast of the center of the remnant. Courtesy NASA/CXC/Rutgers/J.~Hughes et al.
}
\end{center}
\end{figure}

SNR 292.0+1.8 is, along with Cas~A and Puppis, one of three known
oxygen-rich supernova remnants in the Galaxy. Hughes et al.\ (2001)
performed an observation with ACIS-S3 (Figure~\ref{f:SNR292.0+1.8}),
detecting a bright, spectrally hard, point source within an apparently
extended region.  This detection suggested the presence of a pulsar
and its pulsar-wind nebula.  Radio observations (Camilo et al.\ 2002b)
then found a 135-ms pulsar in SNR 292.0+1.8 localized to within the
Parkes beam (~14 arcmin FHWM).  The detection by Hughes et al.~(2003b)
of X-ray pulses at the expected period from the compact X-ray star
secured its identification. The X-ray spectrum is modeled with a
simple power law, although, as with Vela, (and many other sources) the
fit to the data is not unique.  From the motions of oxygen-rich
optical knots and the size of the remnant, Ghavamian, Hughes, and
Williams (2005) recently estimated a kinematic age for SNR 292.0+1.8
of 3000--3400 years assuming a distance of 6 kpc. This value is in
good agreement with the pulsar spin-down age of 2900 years.

\subsubsection{3C58\label{sss:3c58}}

\begin{figure} 
\begin{center}
\resizebox{6in}{!}{\rotatebox{0}{\includegraphics{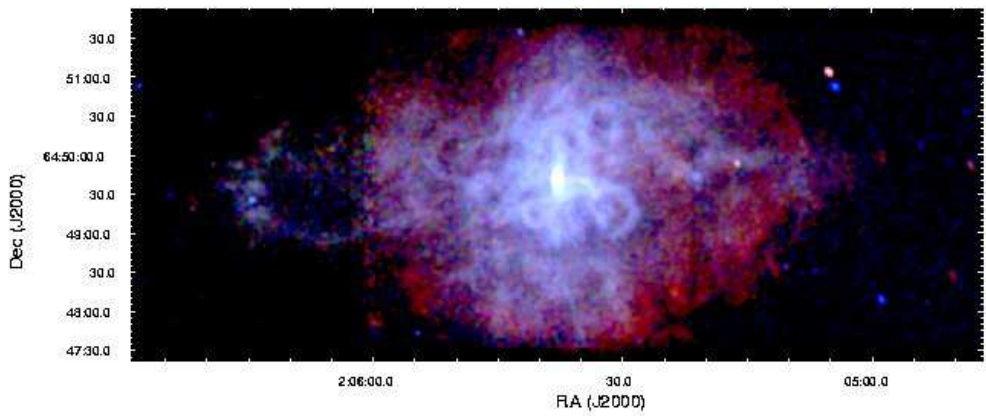}}}
\caption{\label{f:3C58} \cha\ ACIS-S3 image of 3C58. Courtesy P. Slane. 
}
\end{center}
\end{figure}

The \cha\ observations of 3C58 (aka SNR 130.7+3.1) were first
performed by Murray et al.\ (2002b) using the High Resolution Camera
(HRC) which offers excellent time resolution but no spectral
information.  These data imaged the previously detected X-ray point
source (Becker, Helfand, and Szymkowiak 1982) which is associated with
the historical SN 1181 (see Stephenson \& Green 2002 and references
therein).  The early \cha\ data also revealed the extended PWN and the
presence of 66 ms pulsations from the central point source
(J0205+6449).  Deeper \cha\ observations using ACIS-S3 by Slane et
al.\ (2004b) produced images such as that shown in Figure~\ref{f:3C58}
showing the similarity of this PWN with the Crab and Vela.

One aspect of the \cha-based research of 3C58 of special importance
were the limits as to any thermal emission from the surface of this
young cooling neutron star.  The search for thermal emission was
presented by Slane, Helfand, and Murray (2002) and then refined by
Slane et al.\ (2004b) who found that as with SNR 266.2$-$1.2
(\S~\ref{sss:SNR266.2-1.2}) their upper limit ($T_{\infty} < 1.02
\times 10^6$ K) falls well below predictions of standard neutron star
cooling.  Yakovlev et al.\ (2002) discuss calculations of neutron star
cooling in the context of 3C58 and concluded that the observations can
be explained by the cooling of a superfluid neutron star where the
direct Urca process is forbidden.

We note that it is far easier to derive a stringent upper limit to any
thermal component for 3C58, in contrast for example to the Crab
pulsar, because the flux of 3C58 is much lower.  Of course neutron
stars may be different, so that limits to the thermal components of
both sources, indeed all the young neutron stars, are relevant to
compare with theoretical predictions of neutron star cooling.  In
general, such analyses are not simple, requiring enhanced sensitivity
for the detection of the putative thermal component often in the
presence of a much stronger non-thermal flux from the magnetosphere of
the pulsar, if one wants to measure the temperature --- as opposed to
setting an upper limit.  \cha\ is uniquely poised to provide the raw
data for such studies due to its ability to maximally separate the
pulsar from the surrounding nebulosity, yet often long observations are
required.

\subsubsection{IC443\label{sss:ic443}}

\begin{figure} 
\begin{center}
\resizebox{4in}{!}{\rotatebox{0}{\includegraphics{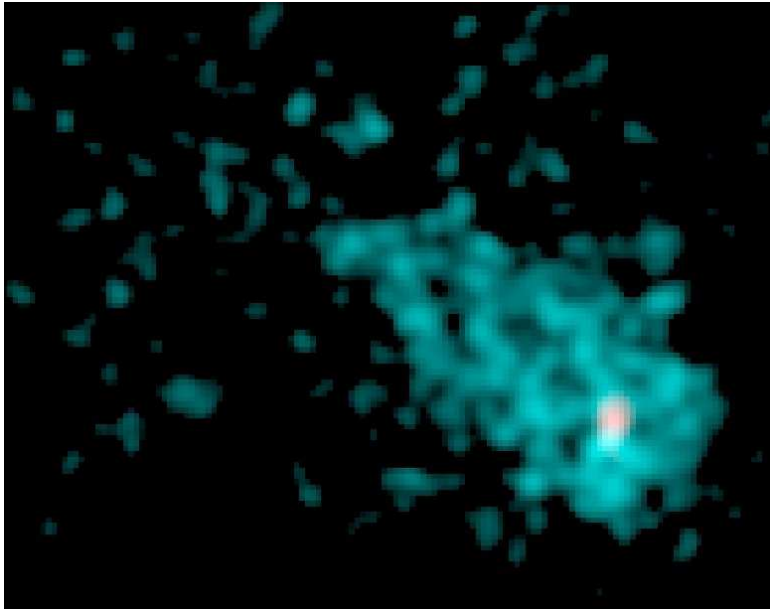}}}
\caption{\label{f:ic443} \cha\ ACIS-I image of IC443. The image is 1.0' x 0.8'. Courtesy  NASA/NCSSM/C. Olbert et al.
}
\end{center}
\end{figure}

Historically, the \cha\ ACIS-I3 image of IC443 shown in
Figure~\ref{f:ic443} was a publicity tour-de-force for the \cha\
project as the first three authors of Olbert et al.\ (2001) were high
school students at the time.  This remnant had been previously well
studied as there is a large variety of shocked molecules present due
to the interaction with surrounding molecular clouds (see
e.g., references in Olbert et al.\ 2001 and Bykov, Bocchino, \& Pavlov
2005).  IC 443 is also a candidate counterpart to the EGRET source,
3EG J0617+2238.  The image (Figure~\ref{f:ic443}) shows what appears
to be a point source behind a bow shock and surrounded by a nebulosity
that looks somewhat like a cometary tail.  Olbert et al.\ (2001) also
reported accompanying VLA observations which confirmed and
complimented the X-ray spatial structure and exhibited varying degrees
of polarization as strong as 25\%.  No pulsations were reported either
from the X-ray or the flat spectrum radio observations.  Subsequent
observations with both \cha\ and \XMM\ (Bocchino and Bykov 2001;
Bykov, Bocchino, Pavlov 2005) have also not detected pulsations.

\subsection{Not finding the compact objects\label{ss:notfindingpulsars}}

In a number of cases high sensitivity searches with \cha\ have been
unable to specifically identify a compact object associated with a
SNR, although often numerous candidate objects have been detected.

\subsubsection{SN 1987A\label{sss:1987a}}

Since launch, SN 1987A has been the focus of a series of repeated
observations with \cha\ (see Park et al.\ 2005 and references
therein).  One goal of these observations is to detect the emergence
of the X-ray flux from a newly born compact object.  To date, no such
object has been detected and Park et al.\ (2005) assume a spectral form
--- a power law of photon index 1.7 --- and use an absorbing column
derived from their fit to the entire remnant to set a 90\% confidence
upper limit to the 3--10 keV luminosity of $1.3 \times 10^{34} $ ergs
s$^{-1}$.  As noted by the authors, the uncertainty as to the correct
column to apply to this calculation could easily increase this upper
limit.

\subsubsection{$\gamma$-Cygni (SNR 78.2+2.1)\label{sss:gammacycgni}}

Becker et al.\ (2004) used \cha\ to search for the X-ray counterpart
to 3EG J2020+4017 (2CG078+2).  In particular these authors were
following up on the possibility (Brazier et al.\ 1996) that RX
J2020.2+4026 was the counterpart.  These observations, thanks to the
precision with which X-ray sources in the field could be located,
demonstrated conclusively that RX J$2020.2+4026$ is associated with a
K field star and therefore an unlikely counterpart of the bright
EGRET source.

This observation also demonstrated the difficulties one sometimes
encounters in searching for compact objects associated with a SNR.
Thus, 37 additional X-ray sources were detected in the field searched
(which was only a fraction of the full size of the SNR).  Radio
observations reported by these authors, which covered the complete 99\% EGRET
likelihood contour of 3EG J2020+4017 with a sensitivity limit of
$L_{820} = 0.09 \mbox{ mJy kpc}^2$, were unable to find a pulsar.  The
absence of radio pulsations suggests that if there is a pulsar
operating in $\gamma$-Cygni, the pulsar's emission geometry is such
that the radio beam does not intersect with the line of sight.
Alternatively, the pulsar is perhaps a CCO-like object which does not
produce significant amounts of radio emission.

Without high-precision X-ray spectra of each of the candidate X-ray
sources, and detailed follow up in other wavelength bands, there is
essentially no satisfactory way in which to eliminate most of the
candidates from consideration.  In such cases, the principal and
important \cha\ contribution is to provide target lists with accurate
positions as a basis for future studies.

\subsubsection{SNRs 315.4$-$2.30, 093.3+6.9, 084.2+0.8, and 
127.1+0.5\label{ss:SNR315.4-2.30}}

Gvaramadze and Vikhlinin (2003) analyzed archival ACIS-I observations
of SNR 315.4$-$2.30 (aka MSH 14$-$6{\sl 3}, RCW 86) a bright, radio
shell-like SNR (see the introduction to Gvaramadze \& Vikhlinin 2003
and references therein; see also \S 5.2 of Kaplan et al.\ 2004 and
references therein).  They concentrated their study to a protrusion in
the southwest of the remnant based on the hypothesis that the SNR
resulted from an off-centered explosion of a moving and massive star.
Two X-ray sources were detected in this region, one of which they
identified with a foreground star.  The second source they identified
as a candidate for the compact remnant, in part because of the
positional coincidence in support of their hypothesis, and in part
because of the absence of an optical counterpart.  The location
implies a transverse velocity of over 1500 km s$^{-1}$.

SNR 315.4$-$2.30 is also one of the four thoroughly-studied SNR by
Kaplan et al.\ (2004). The additional targets are listed in the title
of this section above. Kaplan et al.~have embarked on a program to
perform systematic studies to search for compact central objects in a
distance-limited sample of 23 SNR that lie within 5 kpc.  These
authors include with their \cha\ (and \XMM) observations an
accompanying optical/IR identification program taking account of the
fact that any such counterparts are expected to be very faint.  Using
the brightness of the \cha-discovered CCO in Cas A as a reference,
they find no compact central objects associated with these four SNR to
a limit of 0.1 Cas A or L$_x < 10^{31}$ ergs s$^{-1}$.  We eagerly await
the subsequent papers covering the remainder of the observations.
 
\subsubsection{SNR 41.1$-$0.3 (3C 397)\label{sss:SNR41.1-0.3}}

Safi-Harb et al.\ (2005) used a 66 ksec \cha\ ACIS-S3 exposure to study
3C 397 (SNR 41.1$-$0.3).  One goal of this study was to search the
central X-ray hot spot for a compact remnant left by the original
supernova.  No viable counterpart was found, and these authors placed
an upper limit to the 0.5--10.0 keV flux of $6 \times 10^{-13}$ ergs
cm$^{-2}$ s$^{-1}$ (L$_{x} (0.5-10.0$ keV) $= 7 \times 10^{33} D^{2}$
ergs s$^{-1}$).  Unfortunately there are some ambiguities as to how
these numbers were obtained as the spectrum used and the confidence
levels associated with the uncertainties are not mentioned.
 
\subsubsection{N63A in the LMC\label{sss:n63a}}

Warren, Hughes, and Slane (2003) observed the supernova remnant N63A
in the LMC using ACIS-S3.  No hard X-ray point-source was apparent in
these data and these authors could rule out a young, energetic,
Crab-like pulsar.  They set a $3\sigma$ upper limit to the flux of a
point source assuming a power law spectrum with a number index of $-2$.
The 2.0--8.0 keV flux upper limit was $2.5 \times 10^{-14}$ ergs
cm$^{-2}$ s$^{-1}$, or a luminosity of $7\times 10^{33}$ ergs s$^{-1}$
(assuming a distance to the LMC of 50 kpc) and applied to the region
anywhere in the interior of the SNR.  The luminosity limit for a
source extended over a diameter of 2 pc (appropriate to a pulsar wind
nebula) was higher: $4\times 10^{34}$ ergs s$^{-1}$.

\subsubsection{1E 0102.2$-$72.2 in the SMC\label{sss:e0102nopsr}}

This remnant was discovered during the \eo\ survey of the Small
Magellanic Cloud (SMC) (Seward and Mitchell 1981). Shortly after its
discovery, Dopita, Tuohy, and Mathewson (1981) found oxygen-rich
optical emission from the remnant extended over a diameter of
$\sim$24$^{\prime\prime}$. Subsequent optical spectroscopy (Tuohy and
Dopita 1983) revealed that this material was moving rapidly ($\sim$6500
km s$^{-1}$ FWHM), identifying 1E 0102.2$-$72.2 as the first O-rich
SNR in the SMC. Amy and Ball (1993) suggested that a compact feature
near the remnant's projected geometric center in their high resolution
(3$^{\prime\prime}$) radio image might be ``plerionic'' (i.e.,
emission from a pulsar or pulsar wind nebula). Using a 9 ks ACIS-S3
observation taken during \cha's orbital activation and checkout
period, Gaetz et al.\ (2000) set a 3$\sigma$ upper limit to the
luminosity of a hard power law component (with number index of
$-$2.05) of $\sim$$9\times 10^{33}$ erg s$^{-1}$ (in the energy band
above $\sim$3 keV).

1E 0102.2$-$72.2 is used as a \cha\ calibration target and therefore a
large number of observations are available in the archive.  From a
merged data set of on-axis ACIS-S3 observations with a total exposure
time of 125 ks, we set a 3$\sigma$ count rate limit of
$2\times10^{-4}$ s$^{-1}$ (3-8 keV band) on the X-ray emission near
the center of the SNR at the location of the central radio feature.
For a Crab-like spectrum this rate corresponds to an unabsorbed flux
of $7\times10^{-15}$ erg cm$^{-2}$ s$^{-1}$ (2--8 keV) or an X-ray
luminosity of $3\times10^{33}$ erg s$^{-1}$ for a distance of 60 kpc.

\subsection{Miscellaneous\label{ss:misc}}

\subsubsection{Kes 73 and 1E 1841$-$045\label{sss:kes73}}

The anomalous X-ray pulsar 1E 1841$-$045 is associated with the SNR Kes
73.  See the introduction to Morii et al.\ (2003) and references
therein for an overview of previous observations.  The \cha\
observations by Morii et al.\ (2003) using ACIS-S3 in both
timed-exposure (30 ksec) and continuous-clocking (10 ksec) modes were
the first for which the pulsar could be spatially separated from the
surrounding SNR --- totally for the timed-exposure observation and in
one dimension for the continuous-clocking observation.  The spatial
advantage was partially mitigated by these authors, who chose to
perform all analyses with the continuous-clocking mode data, no doubt
in order to exploit the high time resolution (2.85 ms) for this 11.8 s
pulsar.  The spectral parameters they found from fitting the data to a
power law plus black body were $\Gamma = 2.0 \pm 0.3,~kT = 0.44 \pm
0.02$ keV, and $N_{\rm H} = 2.54 (+0.15,-0.13)$ cm$^{2}$.  We note that the
\cha\ response function provided for the timed-exposure mode is not
precisely transferable to continuous clocking mode data, especially
the gain.  Sophisticated users tend to let the gain be an
additional free parameter when spectrally fitting continuous-clocking
mode data.  A further indication that an incorrect response function
may been applied to these data may be indicated by a rather dramatic
and large residual in the spectral fitting at about 1.6 keV, which the
authors attributed to the aluminum in the ACIS filters.  A similar
feature, and at the same energy, was also seen in continuous-clocking
mode data by Patel et al.\ (2003; their Figure 4) who also applied the
timed-exposure mode response to continuous-clocking mode data.  Since
such a feature should {\em not} be present, given a proper response
function, this coincidence may well indicate that the incorrect
response function was used in both cases.  If so, it is not clear,
however, as to what impact (if any) this might have on the derived
spectral parameters.  Further work is needed to clarify this issue.
It is also possible that the timed-exposure mode response functions
used at the time were simply incorrect, leading to spurious features
near the aluminum edge.  A good check would have been (and is) to
compare the phase-averaged pulsar spectrum determined from the data in
both modes.  Such a comparison would be meaningful if pileup had not
been a problem --- unfortunately in this case it was (Wachter et al.\ 2004).  Even so,
comparing timed-exposure mode and continuous-clocking mode spectra
from sufficiently large regions, and well away from the pulsar, might
have sufficed and would have been informative.

Wachter et al.\ (2004) used the timed-exposure mode data to achieve a
precise position for 1E1841$-$045.  The location enabled them to
accomplish a refined, and successful search for an infrared
counterpart.  The archival data from this observation should be also
be analyzed both to establish the validity of the spectrum deduced by
Morii et al.\ (2003) and for any insights that they might provide
concerning the spectrum of the extended emission.

\subsubsection{SNR 109.1$-$1.0 and 1E 2259+586\label{sss:109.1-1.0}}

We include this \cha\ observation of 1E 2259+586 and SNR 109.1$-$1.0
(aka CTB 109) as the anomalous X-ray pulsar 1E 2259+586 lies along the
line of sight to the SNR and may well be associated with it.  For a
discussion of the AXP-SNR connection see Gaensler et al.\ (2001).
Patel et al.\ (2001) used \cha\ ACIS-S3 to, amongst other things,
determine the most precise X-ray position of 1E 2259+586.  The \cha\
image also shows clear evidence for extended emission, extending from
about $4\arcsec$ to more than $100\arcsec$, which Patel et al.\ (2001)
attributed to the SNR.  The precise position enabled Hulleman et
al.\ (2001) to perform deep optical and near-infrared observations with
Keck and they found a faint (K$_s = 21.7 \pm 0.2$ mag) candidate
counterpart.

Deeper observations, which might serve to establish a possible PWN,
would seem called for.

\subsubsection{N49\label{sss:n49}}

\begin{figure} 
\begin{center}
\resizebox{4in}{!}{\rotatebox{0}{\includegraphics{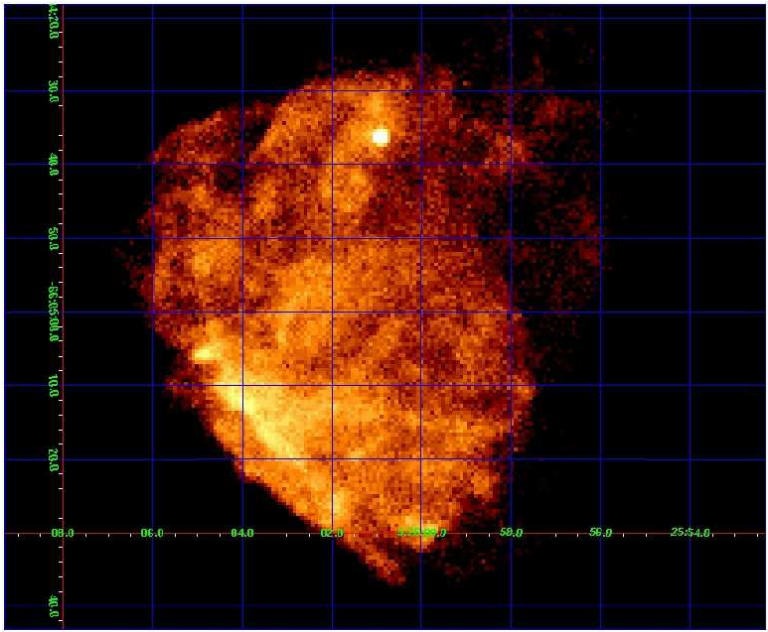}}}
\caption{\label{f:n49} \cha\ ACIS-S3 image of N49. The bright point 
source in the northern part of the remnant is clearly visible. Note
that southern portions of the remnant are cut-off, since the data were
taken in 1/8 subarray mode in order to search for pulsations from  the point 
source. Courtesy S.~Patel.
}
\end{center}
\end{figure}

Park et al.\ (2003) analyzed two ACIS S-3 observations of SNR N49 (aka
SNR 0525$-$66.0) in the LMC.  These authors concentrate their
discussion as to the X-ray properties of the gaseous remnant (see
\S~\ref{jph:n49} below), but do present a brief discussion of the
point source located in the north-eastern portion of the remnant.
This supernova is best known in the high-energy community as the
probable site (Cline et al.\ 1982 --- published in all seriousness on
April 1) of the famous transient event of March 5, 1979, which has
become to be known as the soft gamma-ray repeater (SGR 0526$-$66).
The tie to N49 is as follows.  First, Cline et al. (1982) used data
from a network of satellites, including the \Einstein\ Observatory, to
provide a precise position described by a narrow rectangle which lay
in the northern portion of N49.  The association with N49 was
compelling.  Next, twelve years later, Rothschild, Kulkarni, and
Lingenfelter (1994) using \ROSAT\ discovered an X-ray source whose
positional error circle intersected the error region associated with
the gamma-ray repeater as indicated in their Figure 1.  The \cha\
image of the region containing the \ROSAT\ source is shown in
Figure~\ref{f:n49}.  The \cha\ point source lies within the \ROSAT\
error circle and appears to be within the earlier positional
uncertainty of the SGR.
 
The X-ray source would therefore seem to be the SGR in quiescence, and
using \cha\ ACIS-S3 data Kulkarni et al.~(2003) claim a detection of
8-s periodic pulsations, as were observed during the original
transient gamma-ray event.  The statistical significance of the pulsed
X-ray signal is only modest ($\sim$99.98\% confidence level) and
suggests a spin-down rate of $6.5 \times 10^{-11}$ s s$^{-1}$ based on
two measurements separated by approximately a year and a half.  The
pulse fraction is low $\sim$10\%. The characteristic age of the pulsar
is $\sim$2000 yr, somewhat less than the estimated remnant age of 5000
yr (Vancura et al.~1992).
 
Kulkarni et al.~(2003) and Park et al.\ (2003) also analyzed the
spectral data for the point source.  The spectrum was relatively
featureless, in contrast to the line-dominated spectra from the rest
of the remnant.  Kulkarni et al.~(2003) fit the spectral data with
several models and found that a powerlaw or powerlaw with a blackbody
produced acceptable fits.  The significance associated with inclusion
of the blackbody component, which required a temperature (at infinity)
of 0.5--0.6 keV, was only 90\%.  The photon index of the powerlaw was
$\sim$3.1 and the absorbing column to the point source,
$\sim$$5.5\times 10^{21}$ cm$^{-2}$ is consistent with the absorption
to the SNR.  Park et al.\ (2003) fit the spectral data using power law
models with $\Gamma = 2.8$ to $2.9$ which provided statistically
acceptable fits.  They also noted that a BB model with a temperature
of about 0.5 keV is not ruled out for the low energy data, especially
if one also invokes a hard tail.  The 0.5 keV BB is characteristic of
what one finds for a CCO and it is tempting to speculate as to a
possible tie in.

The inability to distinguish simple spectral models seems to be a
characteristic of the \cha\ and \XMM\ era of CCD-resolution
spectroscopy. In general, the observations are typically much too
short to accomplish such an objective. It is also not clear that more
sophisticated (admittedly unidentified) mathematical tools are needed
to tackle this problem.  For example, can one take advantage of the
fact that the $\chi^2$ statistic used as a measure of goodness-of-fit
is {\em not} distributed as $\chi^2$ when the model spectral
distribution is not representative of the true, underlying, spectrum?

\subsubsection{NGC 6822 (Ho 12)\label{sss:ho12}}

\begin{figure} 
\begin{center}
\resizebox{4in}{!}{\rotatebox{0}{\includegraphics{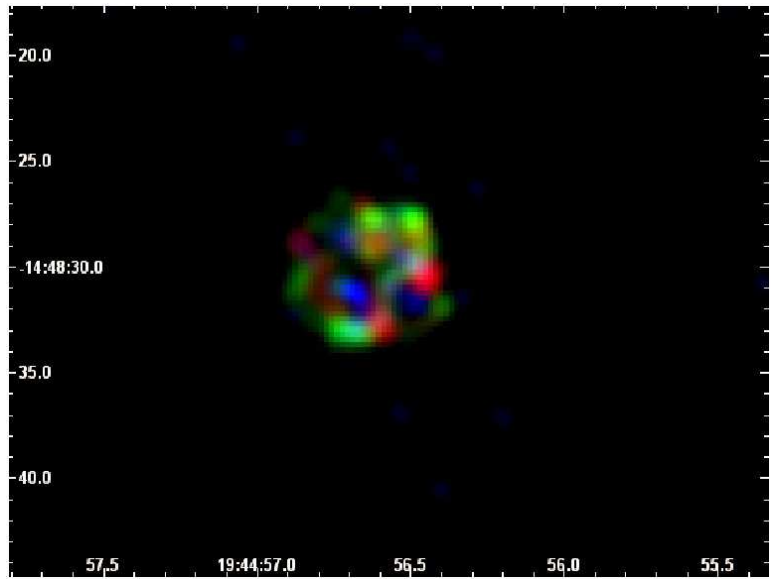}}}
\caption{\label{f:ho12} \cha\ ACIS-I image of Ho 12. Image courtesy D.S. Moon.
}
\end{center}
\end{figure}

Kong, Sjouwerman, and Williams (2004) analyzed archival \cha\ ACIS-I
observations that included SNR Ho 12 in the nearby dwarf irregular
galaxy NGC 6822.  Ho 12 was known to be a SNR based on optical imaging
and spectroscopy and an X-ray source had been associated with the SNR
based on \Einstein\ imaging --- see the introduction to Kong,
Sjouwerman, and Williams (2004) and references therein.  The \cha\
observation (Figure~\ref{f:ho12}) resolves Ho 12 into what appears to
be a shell-shaped object about $10^{\prime\prime}$ (about 24 pc at 500
kpc) in diameter.  The extended image, together with a comparison to
optical and radio observations, unambiguously confirms the
identification with the SNR.  No evidence for, nor an upper limit to,
the presence of a point source was presented.  This object is an
excellent candidate for further study and deeper observations.

\section{Basic View of The Debris and Gaseous Parts of Remnants}

Here we summarize the basic picture of the origin, development, and
evolution of the gaseous remains of supernovae in order to provide
some context for the succeeding discussion.  We will see that \cha\
observations have done much to confirm the basic scenario and also
have revealed new complexities.

As the stellar ejecta stream from the site of the SN explosion, they
expand and interact with the ambient medium (AM), ultimately evolving
into a visible supernova remnant (SNR).  A blast wave, which precedes
the ejecta, forms in the AM, while a reverse shock propagates back
though the ejecta. SNRs gradually become strong X-ray sources over the
course of typically hundreds of years as progressively greater amounts
of ejecta and AM are shock-heated to X-ray temperatures, a consequence
of the high shock velocities (several thousand km s$^{-1}$) during the
early phases of evolution. Typically the X-ray emission is dominated
by the SN ejecta at these early stages, a result of both the higher
density and metal-rich composition of the reverse shocked ejecta.  As
the amount of swept-up AM grows, the ejecta tend to decrease in
importance both in terms of producing X-ray emission and influencing
the remnant's dynamical evolution. According to theory in the
adiabatic phase of evolution, the dynamics depend only on the initial
SN explosion energy and the density of the AM (Taylor 1950; Sedov
1959).  As a remnant ages further its blast wave velocity drops and
radiative cooling at the shock becomes important.  A dense cool shell
of swept up AM forms as all newly shocked material loses its thermal
energy to radiation.  This shell surrounds a hot, low density cavity
of previously shocked ejecta and AM.

The typically low densities and short evolutionary timescales
introduce an important ingredient to the X-ray emission from SNRs,
namely the effect of time-dependent or nonequilibrium ionization.  A
plasma with an electron density of 1 cm$^{-3}$ and temperature $kT\sim
1$ keV takes of order $10^5$ yr to reach collisional ionization
equilibrium. Thus the shocked plasma in remnants, which are almost all
significantly younger than this, will be in a lower state of
ionization than expected based on the plasma temperature.  Departures
from equilibrium ionization are characterized by the ionization
timescale, $n_e t$, which is the product of the plasma electron
density and the time since the material was heated.

The Rankine-Hugoniot (RH) relations (e.g., Shu 1992, p.~214ff) tell us
about the properties of shocks in general, such as how the temperature
and density jump across the shock front depend on the shock velocity,
ratio of specific heats, and other properties of the shocked medium.
However these relations fail to tell us about a number of
astrophysically interesting questions, such as whether individual
species in the fluid (electrons, protons, heavy elements) are heated
to the same temperatures, or how much of the shock energy might be
diverted into a population of relativistic particles (i.e., cosmic
rays). These complications remain open to investigation, even though
the basic properties of SNR shocks are well established.  In the case
of an ideal gas with a ratio of specific heats equal to 5/3, the RH
relations say that a strong shock moving at speed $v_s$ will heat the
gas to a characteristic temperature $kT \sim 5 {\, \rm keV} (v_s/1000
{\, \rm km s^{-1}})^2$.  This clearly corresponds to emission in the
X-ray band, where \cha\ has significant sensitivity.

\section{Studying Ejecta in Supernova Remnants with \cha}

\subsection{Remnants of Young Core Collapse SNe}

The most secure way to identify the remnant of a core collapse SNe is
through the presence of an associated compact object.  In lieu of this
we consider oxygen-rich ejecta to be the likely indicator of a core
collapse SN.  In the following we discuss a selection of remnants that
satisfy these requirements.  Objects are ordered by approximate
chronological age.

\subsubsection{SN 1987A\label{jph:sn87a}}

During the six years covered by this review SN1987A has been monitored
at least yearly by \cha. From Oct 1999 until Jan 2004 the 0.5--2 keV
X-ray flux of the SNR has increased by a factor of 5 and appears to be
growing brighter exponentially (Park et al.~2005).  Even from the
earliest \cha\ observations, deconvolved ACIS images have shown a
ring-like geometry.  That ring is expanding radially at $\sim$4000 km
s$^{-1}$ (Park et al.~2004b). The ACIS-S3 spectra are consistent with
circumstellar matter; there is no evidence yet for X-ray emitting
ejecta in SN198A.

\subsubsection{Cas A\label{jph:casa}}

\begin{figure} 
\begin{center}
\resizebox{4in}{!}{\rotatebox{0}{\includegraphics{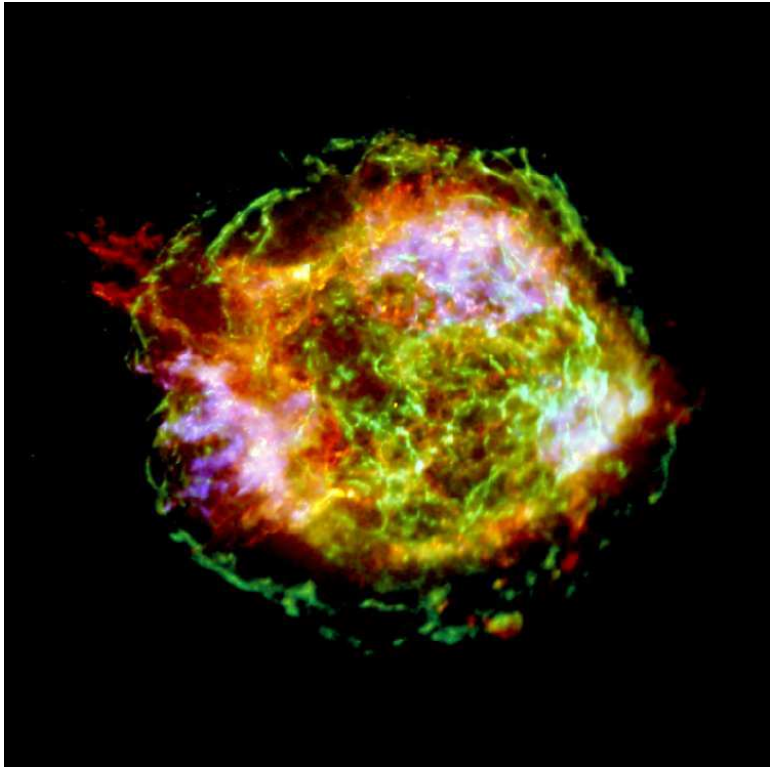}}}
\caption{\label{f:casa_color} Color coded energy image of Cas A using
\cha\ ACIS-S3 data. The image is $6\arcmin \times 6\arcmin$.
The red band contains the Si XIII He$\alpha$ line blend, 
green band the 4.2--6.4 keV continuum, and the blue band contains the
Fe K$\alpha$ line. Image courtesy NASA/CXC/GSFC/U.~Hwang et al.
}
\end{center}
\end{figure}

Cas A (Figure~\ref{f:casa_color}) was the subject of the first
refereed publication on \cha\ observations (Hughes et al.~2000a) and
has remained an intense focus of study through the subsequent years.
This article laid out evidence for three clearly distinct spectral
types in Cas A: Si-rich thermal emission, Fe-rich thermal emission,
and featureless continuum emission.  The Si-rich spectra were
dominated by emission from Si and S with little Fe, with a composition
similar to that expected from explosive oxygen burning.  The Fe-rich
spectra showed strong Fe-L and Fe-K shell emission with much weaker Si
and S lines, consistent with explosive silicon burning.  A further and
perhaps more remarkable finding was that the Fe-rich material along
the eastern edge of Cas A lay outside the Si-rich material, an
inversion of the relative ordering in which these species should have
been produced during the SN explosion.  This argues for an energetic
process of mixing or overturning of the ejecta possibly from
neutrino-driven convection.  The spectrum of the northeast jet was
shown to be less Fe-rich than the knots toward the east.  The
filaments showing featureless continuum emission are possibly due to
synchrotron radiation from relativistic electrons (see
\S\ref{sssjph:crel} below).

Willingale et al.~(2002) used \XMM\ data to measure the radial
velocities of the Si and Fe emission in Cas A.  The southeastern knots
are generally blueshifted while the northern ones are redshifted and
their total velocity range covers $\sim$5000 km s$^{-1}$.  The Si
velocity results are consistent with earlier measurements (Markert et
al.~1983; Hwang et al.~2001) while the results on the velocity of Fe
are new.  The velocities of the southeastern Si and Fe knots are
consistent with each other and confirm that the inversion of the
relative ordering between the Si and Fe emission in this region seen
in the \cha\ images is not a result of some strange velocity structure
in the ejecta.  In addition Willingale et al.~(2002) show that the Fe
emission in the northern region is separated from the Si in terms of
velocity, with the Fe showing higher speeds.  This suggests that the
Si and Fe ejecta experienced a spatial inversion of their original
locations in the north as well.

Laming \& Hwang (2003) and Hwang \& Laming (2003) present a
sophisticated analysis of the X-ray ejecta in Cas A.  Temperature and
ionization timescales determined from spectral fits to individual
X-ray knots were compared to self-similar hydrodynamic models that
incorporate effects of time-dependent ionization, radiative losses,
and the exchange of energy between electrons and ions through Coulomb
collisions. Several sequences of O-rich knots selected from different
parts of the remnant are analyzed in Laming \& Hwang (2003) to infer
variations in the amount of explosion energy directed into different
azimuthal directions.  The asymmetry they determine is as much as a
factor of two with more energy going into the polar regions (near the
jet) than into equatorial regions.  This level of asymmetry is,
however, less than that expected from asymmetric core collapse
explosion models.  Hwang and Laming (2003) study the Fe-rich knots in
Cas A to obtain important constraints on the extent of mixing in the
ejecta.  Several Fe-rich knots along the eastern edge that lie beyond
the Si-rich ejecta are at a mass coordinate of approximately 2
$M_\odot$ measured from the center.  An extremely Fe-rich feature in
the same general vicinity is identified as a possible site of $\alpha$
rich freeze-out (when complete Si-burning occurs at low density).
This material is produced closest to the center of the exploding star
and is therefore most sensitive to the explosion mechanism and the
position of the mass cut between the ejecta and compact remnant.


Gotthelf et al.\ (2001) locate the forward and reverse shocks in Cas
A.  The position of the forward shock was determined by the set of
thin tangential wisps of X-ray continuum emission that extend to a
radius of $153^{\prime\prime} \pm 12^{\prime\prime}$. An increase in
radio intensity as well as a large jump in radio polarization angle
are coincident with the (local) peak in X-ray continuum emission at
this radius. This appears to be the location of the forward shock in
Cas A.  These authors deproject the Si-line image and radio continuum
to estimate the location of the reverse shock, which they determine to
be at a radius of $95^{\prime\prime} \pm 10^{\prime\prime}$. Under the
assumption of standard adiabatic shock models (e.g., Truelove and
McKee 1999), the relative positions of the forward shock and reverse
shock were used to infer that the forward shock in Cas A has swept-up
roughly as much mass as was ejected.  The location of the contact
discontinuity should be determined in order to assess whether the
assumption of standard adiabatic shock models should be revised by,
for example, including the dynamical effects of cosmic ray
acceleration (see \S~\ref{sssjph:crhad} below).

Delaney and Rudnick (2003) determine the average expansion rate of the
forward shock in Cas A to be 4900 km s$^{-1}$ with a range from place
to place that covers $\sim$4000--6000 km s$^{-1}$.  In a subsequent
article Delaney et al.~(2004) carry out an extended study of the
kinematics of Cas A as a function of the spectral character of the
X-ray emission.  They identify four spectrally distinct classes of
emission that they designate as Si-dominated, Fe-dominated,
low-energy--enhanced, and continuum-dominated.  The first two
``ejecta-dominated'' classes show a mean X-ray expansion rate of 0.2\%
yr$^{-1}$, less than the 0.3\% yr$^{-1}$ of the optical ejecta.  This
discrepancy is posited to be due to a greater deceleration of the
lower density X-ray knots compared to the higher density optical
knots.  The low-energy--enhanced component has a low mean expansion
rate 0.05\% yr$^{-1}$ and likely corresponds to the clumpy
circumstellar medium.  Continuum-dominated filaments in the interior
of the remnant also show a low mean expansion rate, although they are
spread over a wide range of expansion rates that include some large
inward motions.  The continuum-dominated filaments around the rim of
Cas A are expanding at the same rate as the Si- and Fe-rich ejecta.

Hwang et al.\ (2004) introduce the Megasec observation of Cas A taken
in 2004 (Figure~\ref{f:casa_color}). These data reveal clearly a
bipolar jet-like structure extending from the northeast to the
southwest.  The jet was mostly likely produced in the SN explosion
process itself rather than being the result of the interaction of the
ejecta with the projenitor's axisymmetric stellar wind.  We expect to
learn much over the next years from this rich data set.

\subsubsection{1E 0102.2$-$72.2\label{jph:e0102}}

\begin{figure} 
\begin{center}
\resizebox{4in}{!}{\rotatebox{0}{\includegraphics{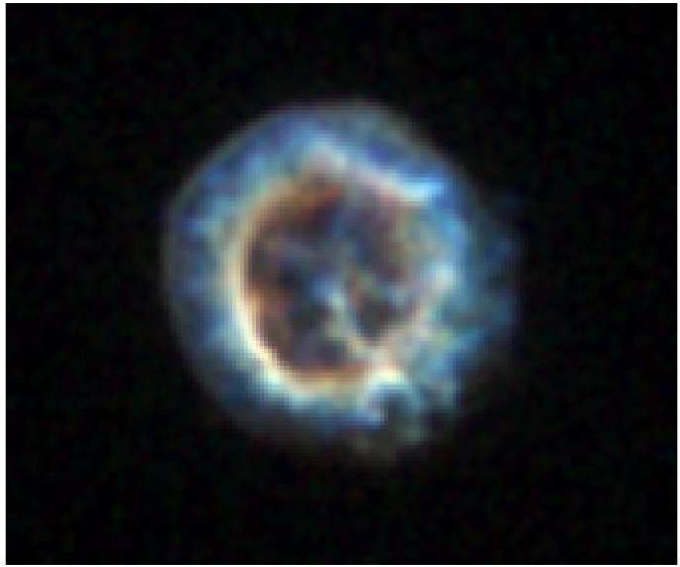}}}
\caption{\label{f:e0102} Color coded energy image of  1E 0102$-$72.2 using
\cha\ ACIS-S3 data. The SNR is roughly 45$^{\prime\prime}$ in diameter.
The red band contains the O VII He$\alpha$ line blend, 
green band the Ne IX He$\alpha$ line blend, and the blue band contains the
Ne X Ly$\alpha$ line as well as Mg and Si lines. 
}
\end{center}
\end{figure}

In the \cha\ ACIS-S3 data of 1E 0102.2$-$72.2 (see Figure~\ref{f:e0102})
Gaetz et al.~(2000) identify a bright, clumpy, incomplete ring of emission
prominent in X-ray lines from He- and H-like oxygen and neon that
correlates well with the optically-emitting ejecta.  Beyond the
ejecta, the \cha\ data show a faint shelf of emission, identified as
the blast wave interacting with the AM and extending to a
diameter of $\sim$44$^{\prime\prime}$.  The rim of the remnant shows a
very sharp, smooth edge, while the outer edge of the ejecta are
``scalloped'' likely as a result of Rayleigh-Taylor instabilities
acting at the contact discontinuity.

Although there is a hint in the ACIS data of spectral variation as a
function of radius through the shock-heated ejecta, this effect is
most clearly seen in the \cha\ High Energy Transmission Grating (HETG)
Spectrometer observation presented by Flanagan et al.~(2004). The size
of 1E 0102.2$-$72.2 is well matched to the dispersion scale of the
grating, so that images of the remnant in individual lines (e.g., O
VIII and Ne X Ly $\alpha$) are cleanly resolved.  Furthermore the
prominence of the remnant's O, Ne and Mg line emission, especially as
compared to Fe, significantly reduces the amount of line-blending in
the 0.7--1 keV band and allows for the derivation of quantitative
results.  Constraints on the global plasma properties were obtained
from measured O and Ne line fluxes (principally the H-like Ly $\alpha$
and He-like resonance and forbidden lines).  The global temperature
values obtained ($kT \sim 0.34$ keV for O and $kT \sim 0.58$ keV for
Ne) are consistent with those from the \XMM\ RGS observation Rasmussen
et al.~(2001).

However, as has been know since the time of \ASCA\ (Hayashi et al. 1994),
the global spectrum of this remnant cannot be adequately described by
a single component plasma model.  The HETG data provide keen insight
into why this so. There are azimuthal variations in the relative
brightnesses of HETG spectral line images that suggest large scale
variations in the plasma properties (for example, the ionization state
toward the north is more advanced than the south).  But perhaps of
greater import is the clear evidence for differences in size of the
various spectral line images.  The sense of these differences is that
lines preferentially produced at lower ionization timescales have
images that are smaller, while higher ionization-timescale lines have
larger images.  This trend is fully compatible with a reverse shock
origin, where the innermost visible ejecta are the most recently
shocked, while the outermost has been shocked for the longest time.
These observational results provide important constraints on the
profiles of temperature and ionization timescale throughout the
shocked ejecta.

There are systematic distortions in the spectral line images between
the + and $-$ arms of the HETG spectrum due to Doppler shifts that
indicate internal velocity differences within the SNR.  The inferred
bulk matter flows are of order 1000 km s$^{-1}$.  Additionally the
ejecta appear to be distributed in a nonspherical manner (toroidal or
cylindrical geometry).  This type of study appears to be one that only
\cha\ is capable of pursuing.

\subsubsection{SNR 292.0+1.8\label{jph:g292}}

The ejecta and shocked ambient medium of SNR 292.0+1.8 were first
studied by Park et al.~(2002) using an ACIS-S3 observation
(\S\ref{sss:SNR292.2}).  The remnant is bright in thermal emission
from O, Ne, and Mg, but weak in Fe, in strong contrast to the
other young Galactic O-rich SNR Cas A. The \cha\ image
(Figure~\ref{f:SNR292.0+1.8}) shows a complex set of knots and
filaments, as well as more diffuse, extended emission. Equivalent
width maps were used to identify metal-rich ejecta, which are
preferentially distributed around the periphery.  A prominent feature
in the broadband image is a narrow, twisted belt of emission running
approximately east-west across the center of the SNR.  It is composed
of solar-abundance material and hence plausibly is shocked
circumstellar gas. Park et al.~(2002) argue that it is the relic of the
axisymmetric dense stellar wind lost by the progenitor star during its
red supergiant phase of evolution.  Another feature noted by these
authors was a set of narrow filaments with normal composition located
around nearly the entire circumference of the SNR. It was suggested
that these filaments might be a relic structure in the circumstellar
(perhaps related to the progenitor's stellar wind) being overrun by
the blast wave.  The more conventional explanation presented by these
authors that the narrow, thermal rims represent the outer blast wave
may be supported by the presence of similar features at the outer edge
of 1E 0102.2$-$72.2 (Figure~\ref{f:e0102}).

In a subsequent study Park et al.~(2004a) examine in more detail the
ACIS-S3 spectra of a number of small-scale emission features that
cover a representative range of spectra from across the SNR.  A
spectrum from the central belt is shown to have abundances that,
within a factor of two, are consistent with solar.  Other regions,
chosen for their prominent line emission, are clearly from SN ejecta
and display greatly enhanced abundances of O, Ne, and Mg compared to
Si, S and Fe. These abundance patterns are similar to those in 1E
0102$-$72.2, while they are quite unlike those in Cas A, which is
dominated by Si- and Fe-rich ejecta.
%
%
Additionally SNR 292.0+1.8 and 1E 0102$-$72.2 did not appear to
undergo the same sort of extensive overturning of their ejecta that
Cas A experienced. The origin for these differences among the class of
O-rich SNRs in not understood.

\subsubsection{N49\label{jph:n49}}

Toward the southwest corner of the SNR N49 Park et al.~(2003a) found a
spectrally hard X-ray knot beyond the main boundary of the
remnant. This feature shows a metal-rich spectrum with apparently
enhanced abundances of O, Mg, Si, S, and perhaps Fe.  Park et
al.~(2003a) argue that it is a fragment of ejecta.  Since this remnant
hosts the famous soft $\gamma$-ray repeater SGR 0526$-$66, insights
into the nature of the progenitor star based on the properties of the
ejecta would be important to obtain.  Unfortunately no other metal
rich ejecta fragments have been reported and the statistical quality
of the knot just described was barely adequate to reveal its
metal-rich nature.  Deeper insights into the nature of the progenitor
will require a considerably longer observation of N49.

\subsubsection{N63A and the Vela Bullets\label{jph:n63a}}

\begin{figure} 
\begin{center}
\resizebox{4in}{!}{\rotatebox{0}{\includegraphics{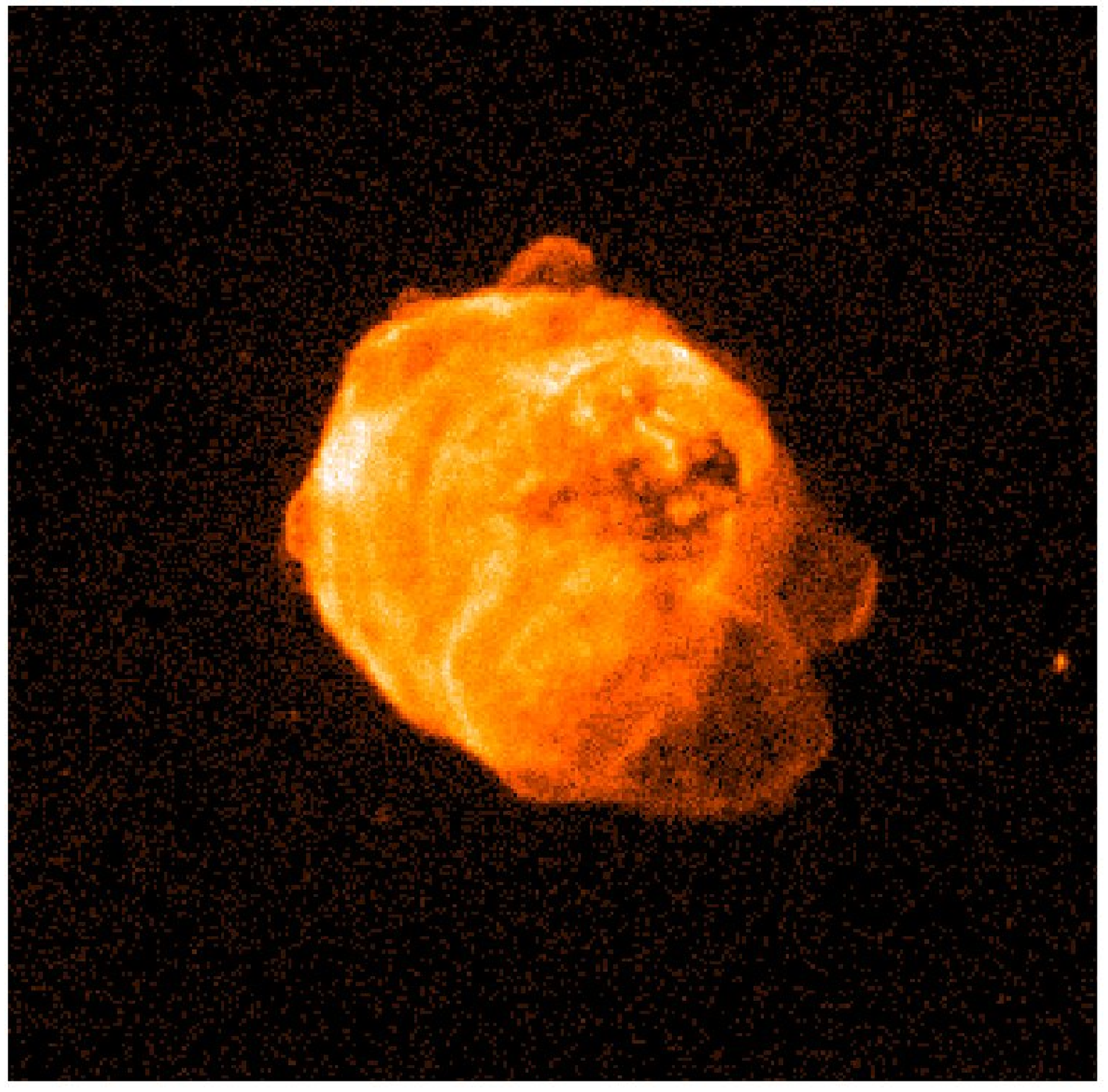}}}
\caption{\label{f:n63a} \cha\ ACIS-S3 image of N63A. The region 
displayed is $150^{\prime\prime} \times 150^{\prime\prime}$.
}
\end{center}
\end{figure}

A dramatic example of an LMC remnant where ejecta fragments, bullets,
or shrapnel may extend beyond the main shock boundary is the case of
N63A (Warren, Hughes, \& Slane 2003).  The crescent-shaped features
seen around the rim here (Figure~\ref{f:n63a}) are very reminiscent of
similar features discovered by \rosat\ surrounding the Vela SNR
(Aschenbach et al.~1995).  In N63A none of the outlying regions shows
strongly enhanced abundances, so if the features are driven by high
speed ejecta clumps, at this point they must have been mixed with
considerable amounts of ambient medium.  Support that the Vela
fragments are in fact SN ejecta have come from \cha\ observations by
Miyata et al.~(2001) of the feature called shrapnel A.  Their ACIS-S3
spectrum from the head of the fragment required twice the solar
abundance of Si.  Shrapnel D is a more definitive case: Katsuda \&
Tsunemi (2005) find factors of 5-10 enhancements in the O, Ne, and Mg
abundances from their \xmm\ observations.  Confidently identifying the
site (or sites) in the exploding star where these fragments originated
based on the observed nucleosynthetic products is a very important
next step.

\subsubsection{SNR 0049$-$73.6\label{jph:0049}}

This object and the one we discuss next are surprisingly similar in
many respects and both appear to be mature versions of 1E
0102.2$-$72. SNR 0049$-$73.6 is 72$^{\prime\prime}$ in diameter
($\sim$21 pc at the distance of the SMC, 60 kpc).  The \cha\ ACIS-S3
data show a central ring-like zone of relatively bright emission
within a faint, circular shell (Hendrick, Reynolds, and Borkowski
2005).  The outer limb-brightened emission is consistent with an
adiabatic shock model with equal electron and ion temperatures
($kT\sim 0.39$ keV) and a preshock hydrogen density of $n_0 \sim 0.14$
cm$^{-3}$.  The remnant's inferred age and explosion energy are 14000
yr and $0.8 \times 10^{51}$ ergs, respectively.  The central bright
emission is enhanced in O and Ne and contains of order 0.35 $M_\odot$
of O, which suggests a progenitor with a mass of at least 13
$M_\odot$.  The authors claim that the observed radius of the ejecta
shell ($\sim$6 pc) is inconsistent with the expected location based on
1D hydrodynamical simulations of a 15 $M_\odot$ progenitor evolving
into a uniform AM.  To explain the apparent discrepancy Hendrick,
Reynolds, and Borkowski (2005) propose the ``Ni bubble'' effect (i.e.,
the dynamical effect of the energy deposited within the center of the
SN by the radioactive decay of$^{56}$ Ni). This is an intriguing idea,
but needs to be developed more fully, in addition to alternative
explanations, such as evolution into a stellar wind density profile.

No mention is made of a compact remnant. We inspected the \cha\ data
and note the presence of a compact source in the hard band ($>$ 3 keV)
about two-thirds of the way out from the center to the outer limb of
the remnant. Although this is likely to be an unrelated extragalactic
background source we use it to set a limit on the presence of an
associated X-ray pulsar.  The 3$\sigma$ count rate upper limit for
this source is $1\times10^{-3}$ s$^{-1}$ (3-8 keV band), which for a
Crab-like spectrum corresponds to an unabsorbed flux of
$3.4\times10^{-14}$ erg cm$^{-2}$ s$^{-1}$.  This yields an X-ray
luminosity upper limit of $1.5\times10^{34}$ erg s$^{-1}$ (2--8 keV
band).

\subsubsection{SNR 0103$-$72.6\label{jph:0103}}

Park et al.~(2003b) observed SNR 0103$-$72.6, the second brightest
X-ray remnant in the SMC, with ACIS-S3.  This observation resolved the
remnant into a nearly complete, remarkably circular shell of emission
85$^{\prime\prime}$ in diameter ($\sim$25 pc at the SMC distance)
surrounding a smaller region with bright clumpy O- and Ne-rich
emission.  The fitted O and Ne abundances of the central ejecta
emission are significantly enhanced above the SMC values, while Fe is
rather low. The relative abundances of O, Ne, Mg, and Si are
reasonably consistent with the nucleosynthetic yields from an 18
$M_\odot$ star.  The outer zone is consistent with blast wave emission
from a remnant in the adiabatic stage of evolution.  The preshock
hydrogen density of the AM is $n_0 \sim 0.16$ cm$^{-3}$, the SN
explosion energy is $1.2 \times 10^{51}$ ergs, and the age of the
remnant is $\sim$18000 yrs.  This is the oldest known example of an
O-rich SNR and was the first one discovered by its X-ray emission.

Park et al.~(2003b) state that all the remnant's emission lies below 3 keV,
but give no quantitative upper limit on the flux of a compact source
with a hard power-law spectrum.  The brightest point source within the
extent of the SNR (according to our own inspection of the \cha\ data)
provides a 3$\sigma$ count rate upper limit of $4\times10^{-4}$
s$^{-1}$ (3-8 keV band), which for a Crab-like spectrum corresponds to
an unabsorbed flux of $1.4\times10^{-14}$ erg cm$^{-2}$ s$^{-1}$ or a
luminosity of $6\times10^{33}$ erg s$^{-1}$ (2-8 keV band).

\subsection{Remnants of Type Ia SNe}

Other than with an optical spectrum from near the time of optical
maximum, there is no definitive way to identify the remnant of a Type
Ia SN (SN Ia). (In a remarkable advance it may become possible to
use light echoes to obtain optical spectra from ancient SN; see 
Rest et al.~2005.)
%
%
In some cases, historical light
curves have been used, although this can only be suggestive. Several
other methods have been proposed based on the expected properties of
the progenitor or its environment, such as (1) a Chandrasekhar mass of
ejecta, (2) partially neutral surrounding ambient medium (Tuohy et
al.~1982), or (3) the X-ray emission from the shocked ejecta (Hughes et
al.~1995).  An additional factor, although not definitive, is the
absence of a compact remnant.  One final introductory note is that,
because the ejecta in SN Ia remnants are essentially invisible in the
optical band, the best way to study it is in the X-ray band.

\subsubsection{Tycho\label{jph:tycho}}

\begin{figure} 
\begin{center}
\resizebox{4in}{!}{\rotatebox{0}{\includegraphics{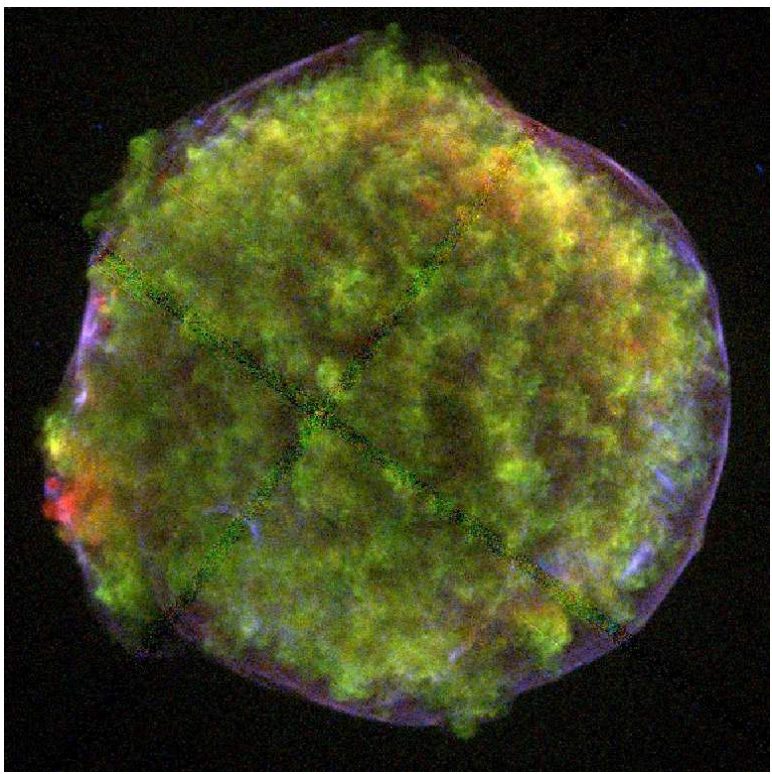}}}
\caption{\label{f:tyc_color} Color coded energy image of the Tycho 
SNR using \cha\ ACIS-S3 data. The image is $9.5\arcmin \times 9.5\arcmin$.
The red band contains the Fe L line blend, the green band the Si 
K$\alpha$ lines, and the blue band contains the
4.1--6.1 keV continuum emission. Image courtesy Warren et al.~(2005)
}
\end{center}
\end{figure}

It was Hamilton, Sarazin, and Szymkowiak (1986b), who presented the
first detailed argument that this bright remnant, observed as a SN by
Tycho Brahe and others in 1572, had a SN Ia origin. They demonstrated
consistency between the global X-ray spectra available at the time and
a model of 1.4 solar masses of layered SN Ia ejecta interacting with a
uniform ambient medium.  Much of the Fe in their model was interior to
the reverse shock and hence unshocked.  Since this study, our
understanding of Tycho's SNR has advanced in several areas.  The
ejecta are clumpy (see Figure~\ref{f:tyc_color}), although nowhere
near as inhomogenous as the core collapse remnants Cas A and SNR
292.0+1.8, and reach to the very edge of the rim in places (Hwang et
al.~2002).  Spectral inhomogeneities throughout the ejecta are
generally modest, with two important exceptions.  There is a radial
variation in the line intensities: Si, S, and Fe-L shell line emission
reach their peak surface brightness at larger radii than does the Fe
K$\alpha$ line emission (Hwang and Gotthelf 1997; Hwang et
al.~2002). Toward the eastern limb are a few knots that bulge out
beyond the rim; the spectral differences between these knots (the
northern one is Si-rich, like the bulk of the ejecta, while the
southern one is Fe-rich) were first noted by Vancura, Gorenstein, and
Hughes (1995).  The origin of these small-scale spectral
inhomogenities is not yet explained, although an interesting idea is
that they may be related to the SN Ia ignition process (Hughes et
al.~2005a).  Based on the analysis of a 150 ksec ACIS-I observation,
Warren et al.~(2005) found the contact discontinuity to be
considerably more structured than the blast wave, indicating
the action of the Rayleigh-Taylor instability there.
%

Badenes et al.~(2003), following firmly in the footsteps of Hamilton,
Sarazin, and Szymkowiak (1986b), have developed a detailed model for
the X-ray emission from remnants of SN Ia. Starting from realistic SN
models that span the range of proposed explosion models, the ejecta
are evolved to the remnant stage with attention to relevant heating,
ionization and thermal effects. Comparison of their model to an \xmm\
spectrum of the Tycho SNR (Decouchelle et al.~2001) suggests the
delayed detonation model as the best choice. Badenes, Borkowski, \&
Bravo (2005) further utilize their model to explain the radial
variation in the X-ray spectrum of Tycho spectra as being due to the
temperature profile through the ejecta.  They require a region of
higher temperature and lower ionization timescale in the Fe-rich zone
near the reverse shock, which can result from a modest amount of
collisionless electron heating at the reverse shock. They also argue
that SN Ia ejecta need to be stratified to some extent which puts
serious constraints on current three-dimensional SN Ia deflagration
explosion models which tend to produce well-mixed ejecta.

\subsubsection{0509$-$67.5\label{jph:snr0509}}

\begin{figure} 
\begin{center}
\resizebox{4in}{!}{\rotatebox{0}{\includegraphics{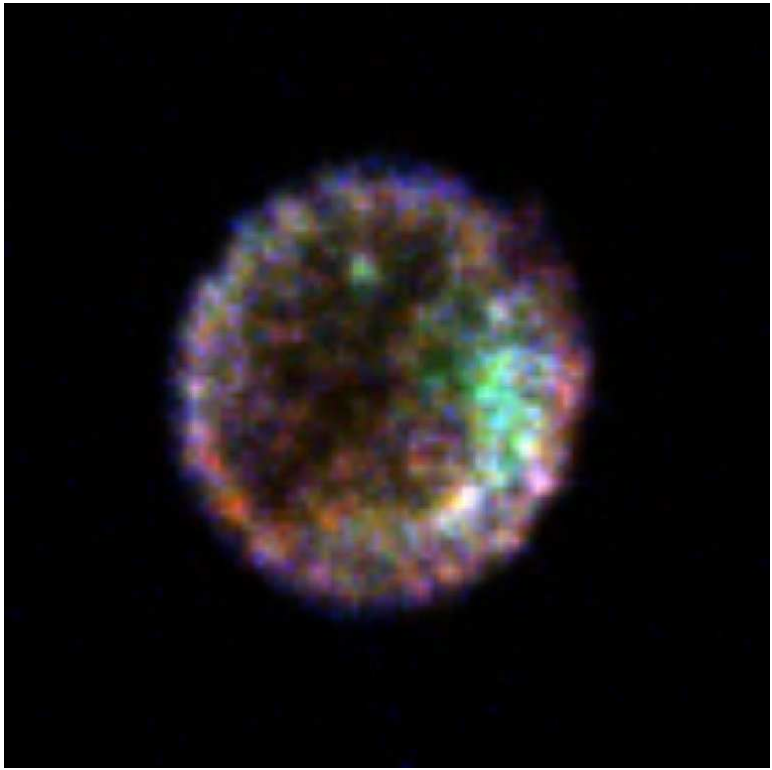}}}
\caption{\label{f:e0509_color} Color coded energy image of SNR
E0509$-$67.5 using \cha\ ACIS-S3 data. The image is $50^{\prime\prime}
\times 50^{\prime\prime}$. The red band contains the 0.2-0.69 keV band, 
the green band the 0.69--1.5 keV and, and the blue band contains the
1.5--7 keV band. Image courtesy Warren and Hughes (2004) 
}
\end{center}
\end{figure}

Like the Tycho SNR this remnant (Figure~\ref{f:e0509_color}) has
provided valuable clues to the SN Ia explosion process. Warren \&
Hughes (2004) show that the integrated ACIS-S spectrum is dominated by
emission from ejecta.  The integrated abundances of O, Ne, Mg, Si, S
Ar and Ca from their fits are consistent with yields from SN Ia
models, with some preference for a delayed detonation model (Iwamoto
et al.~1999).  There is considerably less hot Fe than expected,
arguing that most of the Fe in the ejecta is cold and lies interior to
the reverse shock. The mass of Si inferred from the spectral analysis
is $\sim$0.2 $M_\odot$ when clumping is taken into account, which is
in line with SN Ia models.

\subsubsection{N103B\label{jph:n103b}}

N103B was first suggested to have a SN Ia origin by Hughes et
al.~(1995) based on qualitative inspection of its \ASCA\ X-ray
spectrum.  This was somewhat of a surprise given the remnant's
relative proximity to the young star cluster NGC 1850, only about
40 pc away in projection (Chu \& Kennicutt 1988).  Optically, N103B
consists of several small bright knots that show the usual set of
emission lines seen in radiative shocks: [O {\sc iii}] $\lambda$ 5007,
[S {\sc ii}] $\lambda\lambda$ 6716, 6731, H$\alpha$, and so on
(Danziger \& Leibowitz 1985).  The abundances of N103B inferred from
optical spectroscopy are consistent with the swept-up interstellar
medium of the LMC (Russell \& Dopita 1990).  These characteristics
suggested a massive star progenitor and a dense ambient medium.

In their analysis of the N103B \xmm\ data van der Heyden et al.~(2002)
argue for a massive star progenitor.  They base this conclusion on
abundances derived from RGS data integrated over the entire SNR. They
find more total O compared to Si and Fe than expected from a SN Ia and
suggest a core collapse SN as a viable alternative.  Unfortunately in
their comparison to the model yields they do not include a swept-up
component (which would contribute a significant mass of O) in addition
to the ejecta and so their conclusion on the originating SN type
cannot be considered definitive.

According to Lewis et al.~(2003), who studied the \cha\ ACIS-S3 data,
N103B consists of several spectrally distinct spatial regions. These
include a centrally located region of hot Fe with low ionization
timescale surrounded by a shell-like region dominated by the Si-group
elements, features that resemble those of the Tycho SNR. On the other
hand the O, Ne, Mg, and continuum emission, which are clearly required
by the spectra, do not show a definitive radial trend and, in
particular, are not distributed in a shell-like geometry.  It was
suggested that this emission could be clumpy foreground or background
ambient material overrun by the forward shock. Based on the estimated
masses of the Si-group elements and Fe, Lewis et al.~(2003) argued
that a SN Ia interpretation was preferred.


N103B bears similarities to the Kepler SNR, which also shows an
Fe-rich X-ray spectrum (Cassam-Chena{\"i} et al.~2004), a large
asymmetry in brightness from one hemisphere to the other, more
centrally located Fe emission, and evidence for interaction with a
dense ambient medium. If, at some point, these remnants are
confidently identified with SN Ia events, they might be able to shed
light on the still poorly understood properties of the environment
and/or companion star.

\subsubsection{SN1006\label{jph:sn1006}}

\begin{figure} 
\begin{center}
\resizebox{4in}{!}{\rotatebox{0}{\includegraphics{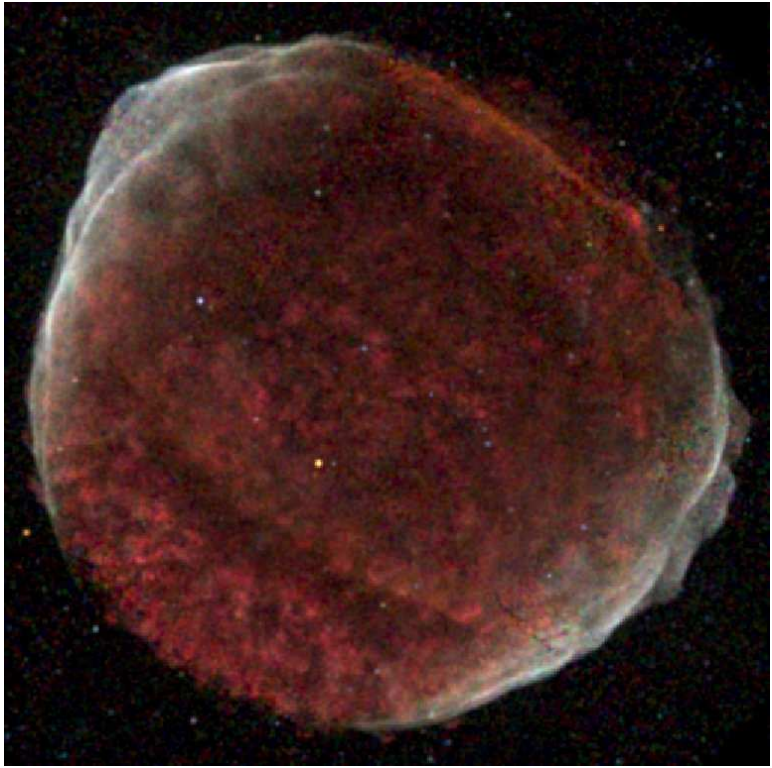}}}
\caption{\label{f:sn1006_color} Color coded energy image of SN1006
using a mosaic of 11 separate \cha\ ACIS-I observations. 
The image is $33.5^{\prime} \times 33.5^{\prime}$. The three
bands (red, green, and blue) correspond to energy bands of 0.50--0.91 keV,
0.91--1.34 keV, and 1.34--3.00 keV.
}
\end{center}
\end{figure}

Hamilton, Sarazin, and Szymkowiak (1986a) attempted to model the
global spectrum of SN1006 in terms of an exploded white dwarf as they
suceeded in doing for the Tycho SNR.  However, unbeknowest to them,
the global spectrum of SN1006 is dominated by nonthermal X-ray
emission coming from the northeast and southwest rims as first
demonstrated by Koyama et al.~(1995) using \asca. These appear as the
whitish rims in Figure~\ref{f:sn1006_color}, which is a complete \cha\
image of SN1006 from a mosaic of eleven individual 22 ksec long ACIS-I
pointings (Hughes et al.~2005b).

The thermal emission from SN1006 is dominated by O emission as shown
by Long et al.~(2003) in their analysis of two deep \cha\ observations
(O emission appears reddish in Figure~\ref{f:sn1006_color}). Toward
the northwest there is a long thin feature, coincident with an
H$\alpha$ optical filament, whose spectrum is consistent with shocked
ambient medium.  Elsewhere the ejecta dominate. SN1006's ejecta
display a fluffy structure that is similar in appearance and physical
scale to the fluff seen in Tycho.  Ejecta clumps in SN1006 also appear
to extend to the forward shock.  ACIS-S3 spectral fits to the ejecta
emission near the northwest filament require enhanced abundances of O,
Mg, Si, and Fe according to Long et al.~(2003). In an analysis of
\xmm\ RGS grating data Vink et al.~(2003) find no evidence for Fe XVII
emission and can fit the EPIC data with emission only from the species N,
O, Ne, Mg, and Si, as well as a hydrogen continuum.

It has been known for some time that SN1006 contains signficant
amounts of cold Fe in its interior based on UV absorption lines in the
continuum emission of a background star (Wu et al.~1983).  This has
been one of the strongest arguments in support of its SN Ia origin.
The presence of Fe in the shocked ejecta is therefore of some
importance, since it would tell us about the extent of radial mixing in
these explosions.

\subsubsection{DEM L71 and other older SN Ia SNRs\label{jph:deml71}}

\begin{figure} 
\begin{center}
\resizebox{4in}{!}{\rotatebox{0}{\includegraphics{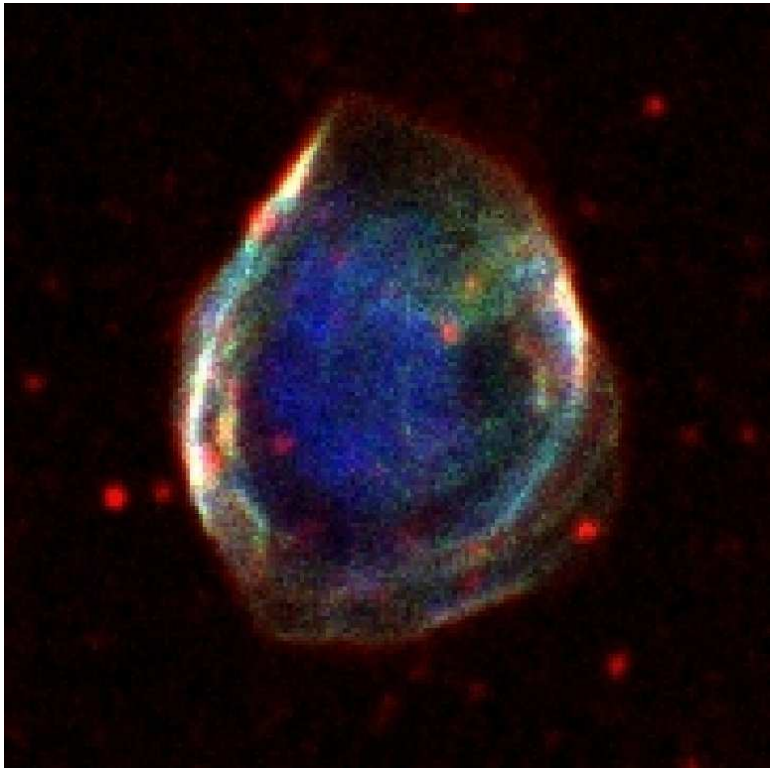}}}
\caption{\label{f:dem71_color} Composite image of DEM L71. 
The image is $2^{\prime} \times 2^{\prime}$.  The red band 
contains H$\alpha$ optical emission while the green and blue bands
include two energy bands from the \cha\ ACIS-S3 data (0.2--0.7 keV and 
0.7--2.6 keV, respectively). Image courtesy Hughes et al.~(2003)
}
\end{center}
\end{figure}

The LMC SNR DEM L71 is some 4000 yr old (Ghavamian et al.~2003).  The
\cha\ ACIS-S3 data (Figure~\ref{f:dem71_color}) show a clear double
shock morphology: an outer rim present in both the broadband X-ray and
H$\alpha$ optical emission and a central excess only seen in the X-ray
band, predominantly above 0.7 keV (Hughes et al.~2003).  The outer rim
is the blast wave in the ambient medium, while the central excess
corresponds to Si and Fe-rich ejecta. The ratio of Fe to O in the
ejecta is at least 5 times the solar ratio.  From the relative
locations of the contact discontinuity and the blast wave shock an
estimate of the total ejecta mass of $\sim$1.5 $M_\odot$ was
obtained. Masses estimated from the X-ray emission are generally
consistent with this value and further show that the mass of Fe is
some 6 times that of Si.  These properties are fully consistent with
the picture that DEM L71 is a middle-aged SN Ia remnant.

It is possible that even older SN Ia remnants may have been
identified.  Hendrick, Borkowski, \& Reynolds (2003) report on the
discovery of metal-rich ejecta in the LMC SNRs 0548$-$70.4 and
0534$-$69.9, which are estimated to be $\sim$10,000 yr old. Their
spectral analysis yields a value for the ratio of O/Fe in the ejecta
that is intermediate between the typical SN Ia and core collapse SN.
On the other hand the mass limits they derive are of order 1 $M_\odot$
or less, supportive of the SN Ia hypothesis.

\section{Shock Properties}

\subsection{Electron-ion temperature equilibration}

For a current review of this topic see Rakowski (2005).  In the
following we highlight studies that utilize X-ray emission to 
investigate the heating of electrons, protons, and heavier ions
at high Mach number shocks.

In a pair of papers on the LMC SNR DEM L71, Rakowski, Ghavamian, and
Hughes (2003) and Ghavamian et al.~(2003) use optical and X-ray
observations to separately measure the post-shock proton and electron
temperatures. The proton temperature comes from the width of the broad
H$\alpha$ line component (for details see Ghavamian et al.~2001 and
references therein) and corresponds to a very thin region
($<$10$^{-3}$ pc) right behind the shock front. The X-ray emission
constrains the electron temperature.  Since even \cha's spatial
resolution is insufficient to resolve the post-shock region, modeling
was required in order to infer the immediate post-shock electron
temperature. Model effects included the temporal evolution in
temperature and ionization timescale as well as equilibration due to
Coulomb collisions between the electron and protons.  Using this
technique it was possible to constrain the initial post-shock ratio of
electron to proton temperatures over a factor of two in shock velocity
(500--1000 km s$^{-1}$).  Lower velocity shocks were found to have
nearly equal post-shock electron and proton temperatures, while the
higher velocity ones showed significantly lower levels of temperature
equilibration.  These results are in good agreement with the published
estimates of the electron to proton temperatures (done with a
different method using only the optical data) from four other SNRs:
SN1006, Tycho, RCW 86, and the Cygnus Loop (Ghavamian et al.~2001).

In their \cha\ HETG study of SN1987A, Michael et al.~(2002)
investigated electron-ion temperature equilibration using X-ray line
widths from a combined line profile including emission from N, O, Ne,
Mg, and Si.  The widths are due to thermal and turbulent broadening as
well as the bulk motion expansion of the remnant.  The first two
effects should produce symmetric, Gaussian profiles, while the latter
profile will depend on the detailed geometry of the expanding
source. Michael et al.~(2002) tried several different geometries for
this and arrived at a shock velocity, $v_s = 3400 \pm 700$ km s$^{-1}$,
that is consistent with the radio and X-ray expansion rates.  Their
measured electron temperature, $kT \sim 2.6$ keV, is much lower than
the post-shock temperature, $\sim$17 keV inferred from the line
widths.  After accounting for Coulomb equilibration between the
electron and ions, the immediate post-shock ratio of electron to shock
temperature can only be $\sim$10\%.
 
Vink et al.~(2003) also use line widths to estimate ion temperatures.
They observed a bright knot along the northwestern limb of SN1006 with
\xmm\ and extracted a spectrum using the RGS.  They found that the O
lines were broadened by $3.4\pm 0.5$ eV, which, if interpreted as a
thermal broadening, indicated an O temperature of $kT = 528 \pm 150$
keV.  The observed electron temperature from this knot was only 1.5
keV, which argues for incomplete electron-ion heating.

Hwang et al.~(2002) observed the forward shock region in the Tycho SNR
with \cha\ ACIS-S3 and set a limit of $<$2 keV on the electron
temperature there.  This was shown to be significantly less than the
shock temperature inferred from the radio (Reynoso et al.~1997) and
X-ray (Hughes 2000) expansion rates.  A limit of 0.03--0.12 was set on
the amount of collisionless electron heating at the forward shock of
Tycho.  We also note that Badenes, Borkowski, \& Bravo (2005) require
roughly the same amount of collisionless electron heating ($<$10\%) at
the reverse shock to explain the X-ray properties of the Fe-rich
ejecta in the Tycho SNR.

\subsection{Cosmic Ray Acceleration}

One of the most important and possibly paradigm-shifting results from
\cha\ observations has come in demonstrating that SNRs are the sites
for the shock acceleration of high energy (TeV range) cosmic rays.
Our discussion of this is divided in two broad categories: evidence
for the electron component and evidence for the hadronic component.
Again we focus on results from the 0.2--10 keV X-ray band.

\subsubsection{Evidence for the Electron Component\label{sssjph:crel}}

Prior to \cha\ there were three shell-like remnants whose emission in
the 0.5-10 keV band was dominated by featureless powerlaw continuum:
SN1006 (Koyama et al.~1995), SNR 347.3$-$0.5 (aka RX J1713.7$-$3946)
(Koyama et al.~1997, Slane et al.~1999), and SNR 266.2$-$1.2 (Slane et
al.~2001).  Other young SNRs (Cas A, Tycho, RCW 86, and Kepler) showed
evidence for hard power-law emission at energies beyond 10 keV (Allen,
Gotthelf, \& Petre 1999).  These featureless power-law spectra, widely
believed to arise from synchrotron radiation, suggested the presence
of highly energetic (TeV range) electrons in these remnants.

\cha\ has contributed to this area in several important ways.  First
was the discovery of geometrically thin, spectrally featureless
filaments in remnants with predominantly thermal emission in the \cha\
X-ray energy band.  Examples include Cas A, where these filaments are
visible as the green-colored network in Figure~\ref{f:casa_color}
(Hughes et al.~2000, Gotthelf et al.~2001), and Tycho, where the thin
bluish rim in Figure~\ref{f:tyc_color} displays a
featureless X-ray spectrum (Hwang et al.~2002; Warren et al.~2005).
Similar thin, spectrally featureless filaments are present in RCW 86
(Rho et al.~2002) and Kepler (Bamba et al.~2005b). In these cases as
well as SN1006 (Long et al.~2003; Bamba et al.~2003), SNR 347.3$-$0.5
(Uchiyama et al.~2003; Lazendic et al.~2004) and SNR 266.2$-$1.2
(Bamba et al.~2005a), the structure of the featureless rims are
consistent with thin sheets of X-ray emission, with widths in most
cases significantly less than a parsec.

Synchrotron cooling of the high energy electrons accelerated at the
shock front is one widely discussed interpretation for the origin of
the thin featureless X-ray filaments (Vink \& Laming 2003; Berezhko,
Ksenofontov, \& V\"olk 2003; Berezhko \& V\"{o}lk 2004; V\"{o}lk,
Berezhko, \& Ksenofontov~2005; Warren et al.~2005).  These studies
provide estimates for the magnetic field within the filaments of order
100 $\mu$G, far above the Galactic value and evidently requiring
magnetic field amplification at the shock front.  An alternate view
(Pohl, Yan, \& Lazarian 2005) posits that the X-ray filaments are
actually magnetic filaments.  In this case the X-ray emission falls
off rapidly behind the shock because the magnetic field, initially
enhanced, is efficiently damped over a post-shock distance of some
$10^{16}$--$10^{17}$ cm.  Careful comparison of radio and X-ray
observations of young remnants might be able to discriminate between
the two interpretations presented here.

\cha\ has also discovered new cases of shell-like SNRs with powerlaw
continuum.  In the case of the ejecta-dominated SNR 0509$-$67.5,
Warren \& Hughes (2004), based on dynamical arguments, propose that
the continuum emission is likely to have a nonthermal origin.  Bamba
et al.~(2004) report the discovery of nonthermal X-ray emission from a
portion of the shell of the LMC superbubble 30 Dor C.  The spectral
index of the emission is consistent with the other cases mentioned
here.  These objects, both in the LMC, are too distant for even \cha\
to provide useful constraints on the thickness of the nonthermal
emission regions.


\subsubsection{Evidence for the Hadronic Component\label{sssjph:crhad}}

Direct detection of the hadronic component of cosmic rays in SNRs is a
critical apect of the overall scenario that remains to be
demonstrated. A prime method for doing this relies on observing TeV
$\gamma$-rays from the decay of pions produced when cosmic ray protons
accelerated at the SNR shock interact with local interstellar
gas. Such emission has not yet been confidently detected.  

If the acceleration process is efficient, a significant fraction of
the SNR shock energy can be diverted from the thermal gas and end up
in the relativistic component. In a study of the SMC SNR 1E
0102.2$-$72.2, Hughes, Rakowski, \& Decourchelle (2000b) measured the
expansion rate (proper motion) of the SNR to determine a blast wave
velocity of $\sim$6000 km s$^{-1}$. From this value they determined a
range of electron temperatures 2.5--45 keV, dependent on the degree of
collisionless electron heating. Their measured electron temperature
from \cha\ ACIS-S3 spectroscopy ($kT \sim 0.4-1$ keV) was significantly
lower than the range expected from the shock velocity.  They argued
that the only plausible way to reconcile the different electron
temperature values was if a significant fraction of the shock energy,
rather than contributing to the heating of the post-shock electrons
and ions, had gone into generating cosmic rays.

It should also be possible to rely on the dynamical effects that
cosmic ray protons have on the evolution of SNRs (Decourchelle,
Ellison, \& Ballet 2000).  The softer equation of state of a
relativistic fluid and the possibility that cosmic ray particles may
escape from the shock result in a compression factor that is larger
than the typical value of 4.  This results in a shrinking of the gap
between the contact discontinuity and forward shock.  Warren et
al.~(2005) measure the relative locations of the forward shock,
contact discontinuity, and reverse shock in the Tycho SNR using \cha\
observations. The observed distance between the contact discontinuity
and forward shock is too small to be explained by hydrodynamic models
that ignore cosmic ray acceleration.  The authors show that the energy
density in relativistic electrons alone is insufficient to alter the
dynamics of the forward shock, which points to the requirement for a
signficant component of relativistic ions. The energy density of the
cosmic ray ions needs to be 50 times or more that of the electrons,
close to the well-established ratio of 100 for Galactic cosmic rays.

\section{Acknowledgments}

Images labeled ``courtesy NASA/...'' are publicly available at
http://chandra.harvard.edu/.  Other contributors of figures are
acknowledged in the figure captions.  MCW appreciates a thorough
reading of, and comments to, a draft of this manuscript by V. Zavlin.
JPH acknowledges C.~Badenes, G.~Cassam-Chena{\"i}, and J.~Warren for
helpful discussions related to this article.


\clearpage

\clearpage

\clearpage

\clearpage

\clearpage

\clearpage

\clearpage

\clearpage

\clearpage

\clearpage

\clearpage

\clearpage

\clearpage

\clearpage

\clearpage

\clearpage

\clearpage

\clearpage

\clearpage

\clearpage

\clearpage

\clearpage

\clearpage

\clearpage

\clearpage

\clearpage

\clearpage

\clearpage

\clearpage


\end{document}